\documentclass{aa}
\usepackage{graphicx}
\usepackage{amsmath}
\usepackage{txfonts}
\usepackage{aalongtable}
 \usepackage{newclude}
\usepackage[round]{natbib}
\bibpunct{(}{)}{;}{a}{}{,}
\graphicspath{{./images/}}
\def\spose#1{\hbox to 0pt{#1\hss}}
\def\lta{\mathrel{\spose{\lower 3pt\hbox{$\mathchar"218$}}
     \raise 2.0pt\hbox{$\mathchar"13C$}}}
\def\gta{\mathrel{\spose{\lower 3pt\hbox{$\mathchar"218$}}
     \raise 2.0pt\hbox{$\mathchar"13E$}}}
\begin{document}

\title{The outer regions of the giant Virgo galaxy M87 \\
   Kinematic separation of stellar halo and intracluster light
  \thanks{Based on observations made with the VLT at Paranal
    Observatory under programs 088.B-0288(A) and 093.B-066(A), and
    with the SUBARU Telescope under program S10A-039.} }

\author{Alessia Longobardi\inst{1}, Magda Arnaboldi\inst{2},
  Ortwin Gerhard\inst{1}, Reinhard Hanuschik\inst{2} }

\offprints{A. Longobardi}

\institute{Max-Planck-Institut f\"ur Extraterrestrische Physik,
  Giessenbachstrasse, D-85741 Garching, Germany \\
  e-mail: alongobardi@mpe.mpg.de, gerhard@mpe.mpg.de \and European
  Southern Observatory, Karl-Schwarzschild-Strasse 2,D-85748 Garching, Germany \\
  e-mail: marnabol@eso.org, rhanusch@eso.org}

\date{A\&A, in press} %Received .......; Accepted .......}

   \authorrunning{A~Longobardi et al.}
   \titlerunning{Distinct stellar halo and intracluster light in the giant Virgo galaxy M87}

% \abstract{}{}{}{}{} 
% 5 {} token are mandatory
 
\abstract 
% context heading ({optional) 
{} 
{We present a spectroscopic study of a sample of 287 planetary nebulas
  (PNs) around the brightest cluster galaxy (BCG) M87 in Virgo A, of
  which 211 are located between 40 kpc and 150 kpc from the galaxy
  centre. With these data we can distinguish the stellar halo from the
  co-spatial intracluster light (ICL) and study both components
  separately.}
{We obtained PN velocities with a high resolution FLAMES/VLT survey
  targeting eight fields in a total area of $\sim$ 0.4 deg$^2$. We
  identified PNs from their narrow and symmetric redshifted
  $\lambda$5007\AA\ [OIII] emission line, the presence of the second
  $\lambda$4959\AA\ [OIII] emission line, and the absence of
  significant continuum. We implement a robust technique to measure
  the halo velocity dispersion from the projected phase-space to
  identify PNs associated with the M87 halo and ICL. Using photometric
  magnitudes, we construct PN luminosity functions (PNLFs), which are complete
  down to m$_{5007}$=28.8.}
{The velocity distribution of the spectroscopically confirmed PNs is
  bimodal, containing a narrow component centred on the systemic
  velocity of the BCG and an off-centred broader component, which we
  identify as halo and ICL, respectively. We find that 243 PNs are
  part of the velocity distribution of the M87 halo, while the
  remaining subsample of 44 PNs are intracluster PNs (ICPNs). Halo and
  ICPNs have different spatial distributions: the number density of
  halo PNs follow the galaxy's surface brightness profile, whereas the
  ICPNs are characterised by a shallower power-law profile,
  I$_{\mathrm{ICL}}\propto R^{\gamma}$ with $\gamma$ in the range
  $[-0.34,-0.04]$. No evidence is found for an asymmetry in the halo
  and ICPN density distributions when the NW and SE fields are studied
  separately. A study of the composite PN number density profile
  confirms the superposition of different PN populations associated
  with the M87 halo and the ICL, characterised by different PN
  specific numbers $\alpha$. We derive
  $\alpha_{\mathrm{halo}}=1.06\times 10^{-8}$
  N$_{\mathrm{PN}}$L$^{-1}_{\odot,\mathrm{bol}}$ and
  $\alpha_{\mathrm{ICL}}=2.72 \times10^{-8}$
  N$_{\mathrm{PN}}$L$^{-1}_{\odot,\mathrm{bol}}$, respectively.  The
  M87 halo PNLF has fewer bright PNs and a steeper slope towards faint
  magnitudes than the ICPNLF, and both are steeper than the standard
  PNLF for the M31 bulge. Moreover, the ICPNLF has a dip at $\sim$
  1-1.5 mag fainter than the bright cut-off, reminiscent of the PNLFs
  of systems with extended star formation history, such as M33 or the
  Magellanic clouds. }
{The BCG halo of M87 and the Virgo ICL are dynamically distinct
  components with different density profiles and velocity
  distributions.  Moreover, the different $\alpha$-parameter values
  and PNLF shapes of the halo and ICL indicate distinct parent stellar
  populations, consistent with the existence of a gradient towards
  bluer colours at large radii. These results reflect the hierarchical
  build-up of the Virgo cluster.}


   \maketitle
% 
% ________________________________________________________________

 \section{Introduction} 

Galaxy halos are faint stellar components made of stars
gravitationally bound to the individual galaxies.  In galaxy clusters
these halos may be surrounded by intracluster stars. The existence of
a diffuse population of intergalactic stars was first proposed by
\citet{zwicky37,zwicky52}. As a consequence of its low surface brightness, it was
only with the advent of CCD photometry that this diffuse stellar
component could be studied in a quantitative way, thus becoming a
topic of interest for observational and theoretical studies.

The formation of intracluster light (ICL) and of the extended halos
around the brightest cluster galaxies (BCG) is closely related to the
morphological transformation of galaxies in clusters. Two of the main
physical processes describing the gravitational interaction between
galaxies during cluster formation and evolution are dynamical friction
\citep{ostriker75,merritt85,taffoni03,boylan08,delucia12} and tidal
stripping
\citep{gallagher72,moore96,GreggWest98,willman04,read2006}. Depending
on its mass, central distance, and orbit, these processes determine the
fate of a cluster galaxy. Dynamical friction is the primary mechanism
dragging a massive satellite towards the host halo’s centre, where it 
merges with the BCG. On the other hand, tidal forces strip stars
from satellite galaxies which end up orbiting the cluster as
unbound ICL \citep{gnedin03,murante04,murante07,mihos04,rudick06}.  In
the ICL and the outer regions of BCGs where the dynamical timescales
are long, fossil records of accretion events can be preserved over
extended periods \citep{willman04,rudick09,cooper14}. Hence, the study
of the luminosity, distribution, and kinematics of galaxy halos and ICL
may provide information on the evolution of galaxies and their host
clusters.

In the last ten years, the analysis of simulated galaxy clusters has
shed light upon the nature and origin of the diffuse stellar component
and its connection with the BCG
\citep{napolitano03,murante04,murante07,willman04,rudick06,rudick09,puchwein10,laporte13,contini14}.
In the framework of cosmological hydrodynamical simulations,
\citet{dolag10} and \citet{cui14} separated stars bound to the cluster
potential from those bound to the BCG by adopting a dynamical
definition of the main galaxy halo and the diffuse light. Tagging
particles as galaxy or intracluster component based on their different
velocity distributions, they identified two distinct stellar
populations in terms of kinematics, spatial distribution, and physical
properties like age and metallicity. Other studies do not adopt this kind of
dynamical definition of the two components and treat the BCG and ICL
as a single system at the centre of the cluster, consisting of all
stars that are not bound to any satellite subhalos in the cluster
\citep[e.g.][]{murante04,cooper14}.

Galaxy halos, diffuse ICL, and their connection with galaxy evolution
in clusters are the subject of many observational studies. Deep
imaging of individual objects
\citep{bernstein95,gonzales00,feldmeier04b,mihos05,krick07,rudick10}
shows that the faint ICL around BCGs often has irregular
morphology, consistent with predictions from simulations.  From
stacking images for a large number of objects
\citep{zibetti05,dsouza14}, average photometric properties were
obtained, showing that the ICL extends to many 100s of kpcs from the
cluster centre.  In the Virgo cluster, \citet{kormendy09} analysed a
sample of ellipticals and spheroidal galaxies and studied their halos
through the light profiles. Comparing their results with the earliest
simulations of \citet{vanalbada82}, they argued that light profiles
with a large Sersic index ($n > 4$) are common in many giant
ellipticals whose origin can be associated with merger processes.

Kinematics and stellar population parameters have only been measured
in a small number of BCG halos, such as (i) NGC 4889 in the Coma cluster
\citep{coccato10}, which shows a change of stellar population at large
radii, (ii) the central galaxy NGC 3311 in Hydra I \citep{ventimiglia10,
  arnaboldi12}, where the kinematics as well as the morphology of the
outer halo signal on-going accretion events, and (iii) NGC 6166 in Abell
2199 \citep{kelson02, bender14}, whose velocity dispersion profile
blends smoothly into the cluster.  In the Coma cluster, the
ICL kinematics suggest an on-going merger of two cluster cores \citep{Gerhard07}.
All these studies aimed to understand the role of tidal disruption
and merger events as the main processes involved in the formation and
evolution of central cluster galaxies and the ICL.

In the nearest clusters, single stars can be used to study the stellar
populations associated with the outer halos and the diffuse stellar
component \citep{ferguson98,durrell02,williams07,yan2008}. Globular
clusters (GCs) have been used to obtain kinematic information in the
outer regions of nearby early-type galaxies
\citep{cote01,schuberth10,strader11,romanowsky12,pota13}.
Planetary Nebulas (PNs) have been targeted in several surveys aimed to
trace the light and motions in galactic halos
\citep{hui93,mendez01,peng04,coccato09,mcneil10,mcNeil12,cortesi13},
the Virgo cluster IC component
\citep{arnaboldi96,arnaboldi03,arnaboldi04,aguerri05,doherty09,castro09,longobardi13},
and the Hydra I and Coma clusters, out to 50-100 Mpc distance
\citep{ventimiglia10,ventimiglia11,gerhard05}.  It was found that the
observed properties of the PN population, such as the $\alpha$
parameter that quantifies the stellar luminosity associated with a
detected PN, and the PN luminosity function (PNLF) correlate with the
age, colour, and metallicity of the parent stellar population
\citep{hui93,ciardullo04, ciardullo10, buzzoni06,longobardi13}.  Thus
PNs can be used to trace these physical quantities of their parent
stellar populations at surface brightnesses too faint for other 
techniques.

The giant elliptical galaxy M87 has one of the oldest stellar
populations in the local Universe \citep{liu05}, and a stellar halo
containing 70\% of the galaxy light down to $\mu_{\mathrm{V}}=$27.0
mag arcsec$^{-2}$ \citep{kormendy09}. It is close to the centre of
sub-cluster A in the Virgo cluster \citep{binggeli87}, the nearest
galaxy cluster, and it is expected to have transformed over larger
timescales because of galaxy mergers \citep{delucia07}. Deep imaging
\citep{mihos05,janowiecki10} has revealed a complex network of faint,
extended tidal features around M87, suggesting that it is not
completely in equilibrium. Thus, M87 and the surrounding Virgo cluster
core are prime targets to address the formation and evolution of
galaxy clusters, ICL, and BCGs. Indeed, M87 is the subject of
many dynamical studies with X-ray measurements
\citep{nulsen95,churazov10}, integrated stellar kinematics
\citep{murphy11,murphy14}, GC kinematics
\citep{cote01,strader11,romanowsky12,zhu14}, and PN kinematics
\citep{arnaboldi04,doherty09}, to estimate its mass and derive the
dark matter distribution.  Using PN kinematics, \citet{doherty09}
identified M87 halo and IC PNs and showed the coexistence, at radii
$>60$ kpc, of a stellar halo bound to the galaxy potential and a
surrounding unbound Virgo ICL.

In this work, we report the results of a wide and high resolution
spectroscopic survey covering the outer regions of M87 out to a
distance of 150 kpc from the galaxy centre.  The aim of this project
is to investigate the halo-ICL dichotomy, making use of a large
spectroscopic sample of PNs (approximately 15 times larger than
the previous sample of \citealt{doherty09}). The paper is structured as
follows: in Sect.~\ref{sec2} we describe the spectroscopic survey
together with the data reduction procedures and the classification of
PN spectra. In Sect.~\ref{sec3} we study the PN phase-space
distribution and dynamically separate the halo and IC populations.
Spatial density distributions are derived in Sect.~\ref{sec4}, and in
Sect.~\ref{sec5} we present the properties of the halo and IC PN
populations in terms of their $\alpha-$parameters and the morphology
of their PNLFs. Finally, we discuss our results in Sect.~\ref{sec6} and
give our conclusions in Sect.~\ref{sec7}.

We adopt a distance modulus of 30.8 for M87
\citep{ciardullo02,longobardi13}, implying a physical scale of 73 pc
arcsec$^{-1}$.

 \section{The FLAMES M87 PN  survey}
\label{sec2}
\subsection{Photometric sample}
\label{sec2.1}
The photometric candidates targeted by our spectroscopic survey come
from an earlier imaging survey \citep{longobardi13}, covering a 0.43
deg$^{2}$ region centred on M87. Images were taken through a narrowband
filter centred on the redshifted [OIII]$\lambda$5007\AA\
emission line at the Virgo cluster distance (on-band image), and
through a broadband V-filter (off-band image).  Because of their
bright [OIII]$\lambda$5007\AA\ emission, extragalactic PNs can be
identified as unresolved emission sources with positive flux on the
on-band [OIII] image and no detection on the off-band image.

We obtained spectra in two observing campaigns.  For the first
spectroscopic campaign, we selected emission line candidates as
objects with positive flux on the colour [OIII]-V band
image\footnote{Before the subtraction the continuum off-band image was
  scaled to the on-band image by a multiplicative scaling factor found
  measuring fluxes from several bright, isolated stars on both
  images.} (hereafter difference method, see
\citet{feldmeier03} for more details). The visual catalogue
extracted on the basis of the difference method consisted of 1074
objects, which covered a magnitude range $23 \le m_{5007} \le29.8$, and
is statistically complete down to $m_{5007} \simeq 28.8$\footnote{The
  transformation between the AB and 5007 magnitudes for the
  photometric narrowband filter is given by $m_{5007}=m_{\mathrm{AB}}
  +2.49$ \citep{longobardi13}.}.

For the second spectroscopic campaign, we carried out a more stringent
selection procedure described in \citet{longobardi13}. In this
procedure, the PN candidates were selected using automatic selection
criteria, based on the distribution of the detected sources in the
colour-magnitude (CM) diagram and the properties of their point-spread
function \citep[PSF; for more details see][]{arnaboldi02,
  longobardi13}. This automatic catalogue is complete within
the magnitude range $26.3 \le m_{5007} \le28.4$.

The combined total input sample for the spectroscopic survey
(visual catalogue plus automatic catalogue)
consisted of 1484 emission line candidates.
 
\subsection{Observations and data reduction} 
\label{section2.2}

\begin{figure}[h] \centering
   \includegraphics[width=10.cm, clip=true, trim=4cm 0.5cm 2.cm 1.5cm]{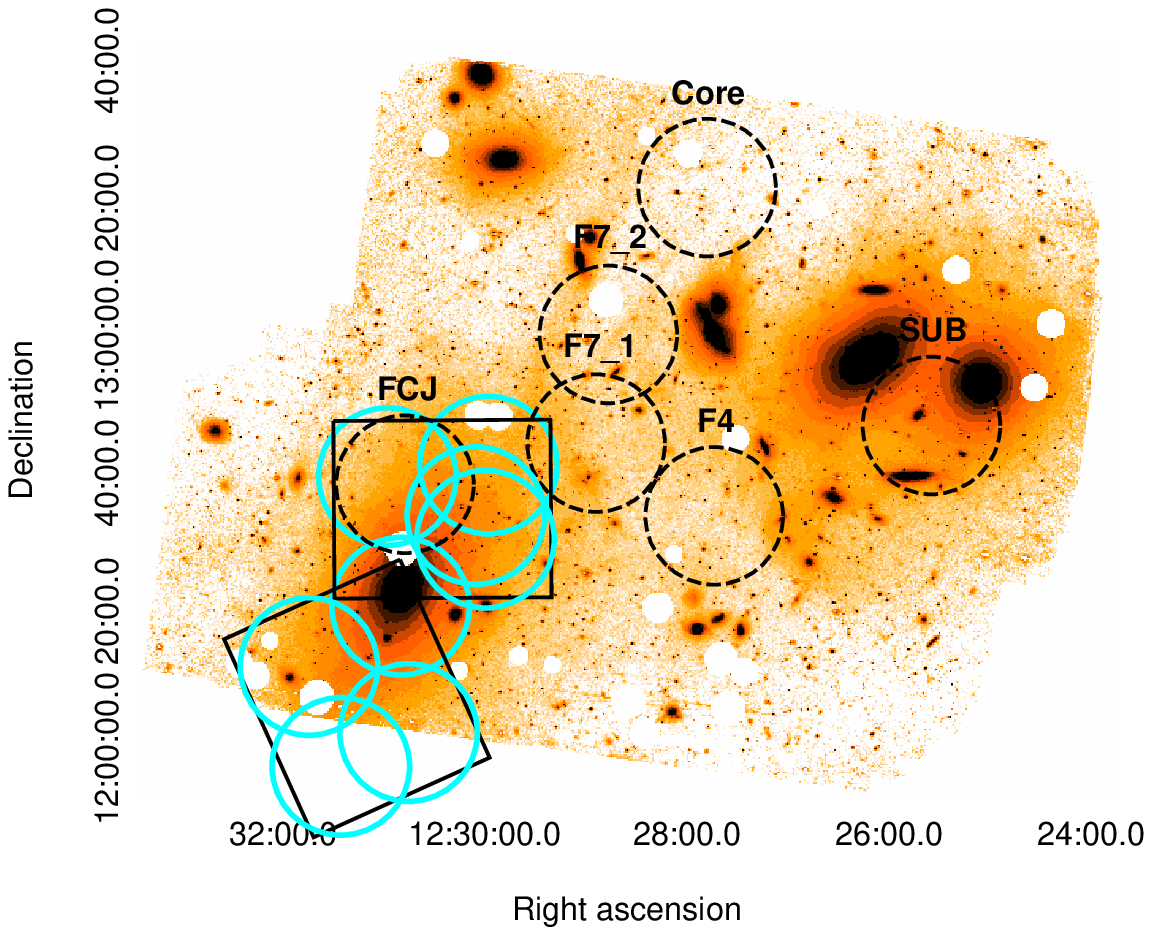}
   \includegraphics[width=8.cm]{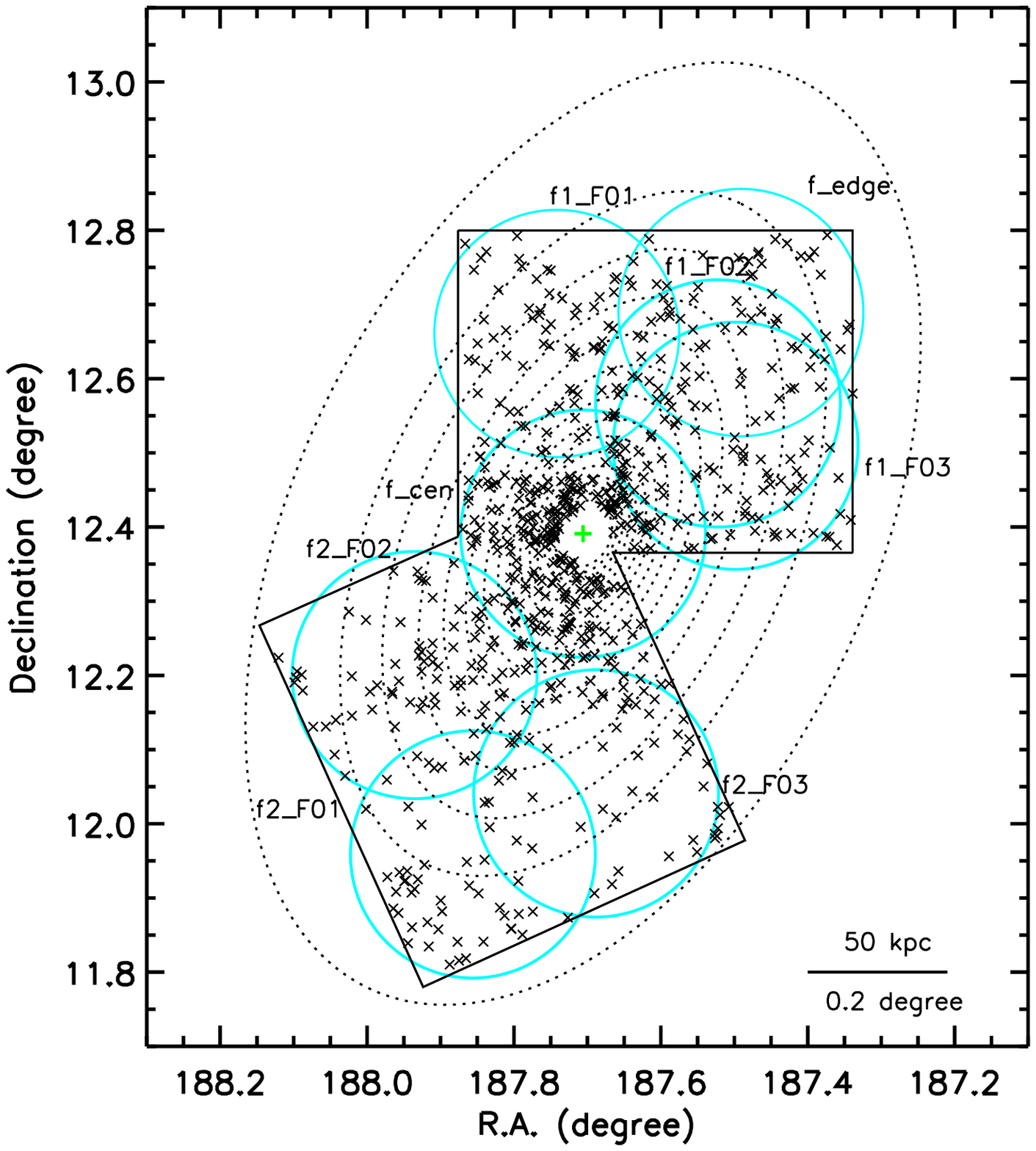}
   
   \caption{\small{\textit{Top panel}: Image of the core of the Virgo
       cluster \citep{mihos05} with the positions of the FLAMES fields
       studied in this work (cyan circles) and in previous surveys (dotted black
       circles) by
       \citet{arnaboldi04} and \citet{doherty09}.  Black squares represent the Suprime-Cam survey
       fields used for the extraction of the photometric PN candidates
       \citep{longobardi13}. \textit{Bottom panel}: Schematic zoom-in
       of the surveyed area (black rectangles). FLAMES pointings (cyan
       circles) and PN candidates from the photometric survey
       \citep[black asterisks;][]{longobardi13} are plotted over the
       isophotes of the M87 halo surface brightness
       \citep{kormendy09}}. The green cross depicts M87's
       centre. North is up, East is to the
     left.}\label{FLAMES_configuration}
 \end{figure}

\begin{table*}
  \caption{\small{Flames configuration and exposure times. 
      Column 1: ID of FLAMES plate. Columns 2 \& 3: FLAMES plate; right ascension 
      and declination. Column 4: Total exposure time. Column 5: Single exposure 
      time $\times$ number of exposures}.}
 \label{F_conf}
   % \centering
  % \caption{\small{Flames Configuration and Exposure Time}}
   \centering
     \begin{tabular}[h]{ c  c  c  c  c c}
     \hline
     \hline
     FLAMES Conf & RA & DEC & Total Exposure time & Single Exposure time\\
     &  &  &  & $\times$ \# of exposures\\
     & h:m:s & degrees & (s)  & (s)\\
     \hline
     M87SUB1 Bright F01 & 12:30:58.015  & +12:39:39.28 &  2700 & 1350 $\times$ 2                \\
     M87SUB1 Bright F02 & 12:30:05.332  & +12:34:00.08 &  2700 & 1350 $\times$ 2             \\
     M87SUB1 Bright F03 & 12:29:59.574  & +12:30:34.49 &  2700 & 1350 $\times$ 2              \\
     M87SUB1 Norm   F01 & 12:30:58.015  & +12:39:39.28 &  8100 & 2700 $\times$ 3 \\
     M87SUB1 Norm   F02 & 12:30:05.332  & +12:34:00.08 &  8400 & 2800 $\times$ 3 \\
     M87SUB1 Norm   F03 & 12:29:59.574  & +12:30:34.49 &  8400 & 2800 $\times$ 3 \\
     M87SUB2 Bright F01 & 12:31:25.426  & +11:57:31.43 &  2700 & 1350 $\times$ 2              \\
     M87SUB2 Bright F02 & 12:31:44.477  & +12:12:01.55 &  2700 & 1350 $\times$ 2             \\
     M87SUB2 Bright F03 & 12:30:45.170  & +12:02:27.20 &  2700 & 1350 $\times$ 2              \\
     M87SUB2 Norm   F01 & 12:31:25.426  & +11:57:31.43 &  8400 & 2800 $\times$ 3 \\
     M87SUB2 Norm   F02 & 12:31:44.477  & +12:12:01.55 &  8400 & 2800 $\times$ 3 \\
     M87SUB2 Norm   F03 & 12:30:45.170  & +12:02:27.20 &  8400 & 2800 $\times$ 3 \\
     M87SUB FEDGE       & 12:29:57.854  & +12:41:20.90 &  6800 & 1700 $\times$ 4 \\
     M87SUB FCEN        &  12:30:50.019 & +12:20:48.88 &   6800& 1700 $\times$ 4 \\
     \hline
         
   \end{tabular}
\end{table*}

The spectra were acquired in service mode with the FLAMES spectrograph
on the VLT-UT2 telescope, in the GIRAFFE+MEDUSA mode. This observing
mode allows for up to 132 separate fibres that can be allocated to targets
in one plate configuration, covering a circular area of 20$\arcmin$
diameter. The total emission line sample was observed in two observing
runs (24h, 088.B-0288(A); 11h, 093.B-0066(A); PI: M.Arnaboldi), which
were characterised by clear conditions and seeing better than
0.9$\arcsec$. We used the high resolution grism HR08, covering a
wavelength range of $\sim250$ \AA\ centred on 5048 \AA\ with a
spectral resolution of $R=22 500$. With this setup, the instrumental
broadening of the arc lines has a FWHM of $17$ km s$^{-1}$ and the
statistical error on the wavelength measurements is 150 m s$^{-1}$
\citep[see][]{royer02}. We refer to Sect.~\ref{vacc} for discussion of
the velocity accuracy estimated from repeat observations of the same
emission line candidates.

Because the [OIII] emission lines from PNs are only a few km s$^{-1}$
wide, high resolution spectra are also desirable to reduce the sky
contamination, making the FLAMES spectrograph the ideal instrument for
LOS velocity measurements of extragalactic PNs.

For our first spectroscopic campaign, the visual catalogue
was divided into a bright ($m_{\mathrm{5007}}< 27.2$) and a normal
($m_{\mathrm{5007}}> 27.2$) sample. The FLAMES plate configurations
and exposure times were then optimised to reach the maximum
number of fibres allocated, as well as optimal signal-to-noise ratio
(S/N) for both samples. Based on the FLAMES/GIRAFFE/HR08
configuration, for a $10^{4}$ sec exposure, the monochromatic
[OIII]$\lambda 5007 \AA\ $ emission of a PN with [OIII] flux of
F$_{5007}=1.0\times 10^{-17} \rm{erg\, cm^{-2}\, s^{-1}}$
(~$m_{5007}~=~28.8$~) is detected with S/N=10 per resolution element
(0.29 \AA).

In total, we defined 12 FLAMES plate configurations, labelled as
M87SUB1 Bright F01-F03, M87SUB1 Norm F01-F03 and M87SUB2 Bright
F01-F03, M87SUB2 Norm F01-F03 for the NW (SUB1) and SE (SUB2)
Suprime-Cam fields, respectively. The layout of the FLAMES pointings on
the sky, together with the coverage of the photometric Suprime-Cam
survey, is shown in Fig.~\ref{FLAMES_configuration}.

In the second spectroscopic campaign, we added two FLAMES plate
configurations covering the very central region of M87 and the NW edge
of the Suprime-Cam imaging survey, in addition to completing the
observations of the 12 FLAMES configurations from the first
campaign. These two additional FLAMES configurations are shown in
Fig.~\ref{FLAMES_configuration} with the labels FCEN and FEDGE,
respectively. Table~\ref{F_conf} provides an overview of the FLAMES
field configurations and the total exposure times.

We reduced the spectroscopic data using the GIRAFFE
pipeline\footnote{https://www.eso.org/sci/software/pipelines/giraffe/giraf-pipe-recipes.html}.
The reduction procedure included bias subtraction, flat-fielding,
identification of the fibre locations on the CCD, geometric distortion
correction, wavelength calibration and extraction of the
one-dimensional spectra. The calibrated one-dimensional
  spectra were then corrected to the heliocentric velocity using the
IRAF\footnote{IRAF is distributed by the National Optical Astronomy
  Observatory, which is operated by the Association of Universities
  for Research in Astronomy (AURA) under cooperative agreement with
  the National Science Foundation.}  task \textit{dopcor}. Finally, we
combined spectra from single exposures using the IRAF task
\textit{scombine} to get the targeted S/N for each spectrum (see
above).

\subsection{Spectroscopic success rates } 
\label{section2.3} 

We define the nominal success rate as the ratio between the number of
spectra with a detected emission line and the number of fibres allocated
for a given FLAMES plate configuration. For our observations, it
varies from field to field, in a range of values between $\sim20\%$
and $\sim60\%$. These values are similar to those obtained in the
spectroscopic follow-up of GCs in the outer halo of M87 (see Sect.~3
in \citet{strader11})\footnote{In \citet{strader11} the
    'nominal' success rate is defined as the ratio between number of
    slit/fibres allocated to candidates and the number of identified
    GCs.}. The low success rates for some of the fields (mostly
M87SUB2 fields) are caused by guide star proper motions, which were not
correctly accounted for in the FLAMES astrometry input file for the
fibre allocation.  We were also able to estimate the fraction of
fibre-object misalignments, from the repeat observations of the
emission objects in common between adjacent FLAMES plate
configurations.  From this we determined the spectroscopic
  completeness C$_{\mathrm{spec,fb}}$. The success rates are affected
by the spectroscopic completeness as well as by the presence of stars
in the catalogue of candidates to which the fibres are allocated.  In
Table~\ref{objfrac}, we report the total number of allocated fibres
together with the nominal success rate, the number of
spectroscopically confirmed PNs, and the spectroscopic completeness
C$_{\mathrm{spec,fb}}$ for all of the FLAMES fields.

Because success rates are also dependent on the absence of
  stars in the submitted catalogue, candidates from the more
  stringent automatic sample have higher success rates than
candidates from the visual catalogue, reaching $>70\%$ in
M87SUB1 BrightF03 and M87SUB2 BrightF01 (see Table~\ref{objfrac}). On
the other hand, about 30\% of the confirmed spectra come from targets
in the visual catalogue: this is consistent with the results
of the simulations by \citet{longobardi13} for the fraction of missed
true line emission sources with automatic selection criteria.
 
 \begin{table*}

   \centering
   \caption{\small{Summary of the total number of allocated fibres, 
       nominal success rate, detected number of PNs, and spectroscopic 
       completeness in all fields. Numbers in brackets refer to 
       sources in common with the automatic sample.}}
   \label{objfrac} 
     \begin{tabular}{ c  c  c  c c}
      
     \hline
     \hline
\\

     FLAMES & \# of targets with& Nominal success rate&  Confirmed PNs  &{\bf C$_{\mathrm{spec,fb}}$}\\
     Conf.&  fibres allocated&  & &\\
          \hline\\

          M87SUB1 Bright F01 & 33\  \  (12)       &52\%   (42\%)    & 8  (3) & 0.6 \\
          M87SUB1 Bright F02 & 41\  \  (19)       &23\%   (32\%)    & 4   (4) &  0.5 \\
          M87SUB1 Bright F03 & 33\  \  (14)       &42\%   (71\%)    & 8   (8) & 0.9 \\
          M87SUB1 Norm   F01 & 125 (58)       &47\%   (57\%)    & 49   (26) & 0.9 \\
          M87SUB1 Norm   F02 & 130 (65)       &52\%   (62\%)    & 55    (35) & 0.9 \\
          M87SUB1 Norm   F03 & 127 (56)       &53\%   (64\%)    & 54    (31) & 0.9 \\
          M87SUB2 Bright F01 & 23\  \  (8)\  \        &61\%   (75\%)    & 5   (3) & 0.8 \\
          M87SUB2 Bright F02 & 31\  \  (16)       &26\%   (25\%)    & 5   (4) & 0.6 \\
          M87SUB2 Bright F03 & 26\  \  (7)\  \        &54\%   (57\%)    & 5   (3) & 0.8 \\
          M87SUB2 Norm   F01 & 104 (40)       &16\%   (17\%)    & 10   (6) & 0.3 \\
          M87SUB2 Norm   F02$^{a}$ & 144 (71)  &42\%   (51\%)    & 44  (27) & 0.6 \\
          M87SUB2 Norm   F03 & 117 (50)       &24\%   (30\%)    & 23   (14) & 0.4 \\
          M87SUB  CEN        & 130 (94)       &54\%   (60\%)    & 63   (51) & 0.9 \\
          M87SUB  FEDGE      & 131 (60)       &27\%   (30\%)    & 30   (16) & 0.8 \\
     \hline
         
   \end{tabular}

   \tablefoot{Fibre configuration was modified between different exposures.}

 \end{table*}

\subsection{Classification of the extracted spectra}
\label{section2.4}
\begin{figure}[h] \centering
   \includegraphics[width=9.cm]{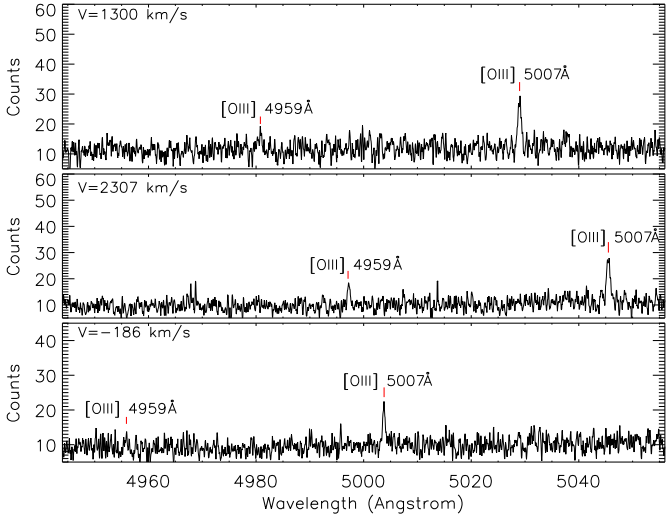}
   \caption{\small{Spectra of single confirmed PNs. The top panel shows the
       spectrum for a PN dynamically bound to the halo component. The
       middle and bottom panels show spectra for PNs dynamically
       unbound to the halo (see Sect.\ref{sec3}), with higher (middle
       panel) and lower (bottom panel) velocity than the M87 systemic
       velocity. Red vertical lines mark the positions of the two
       oxygen lines at their redshifted wavelengths. We smoothed the spectra
       to 0.015 nm per pixel.}}
\label{spectra_example_PN}
 \end{figure}

\begin{figure}[h] \centering
   \includegraphics[width=9.cm]{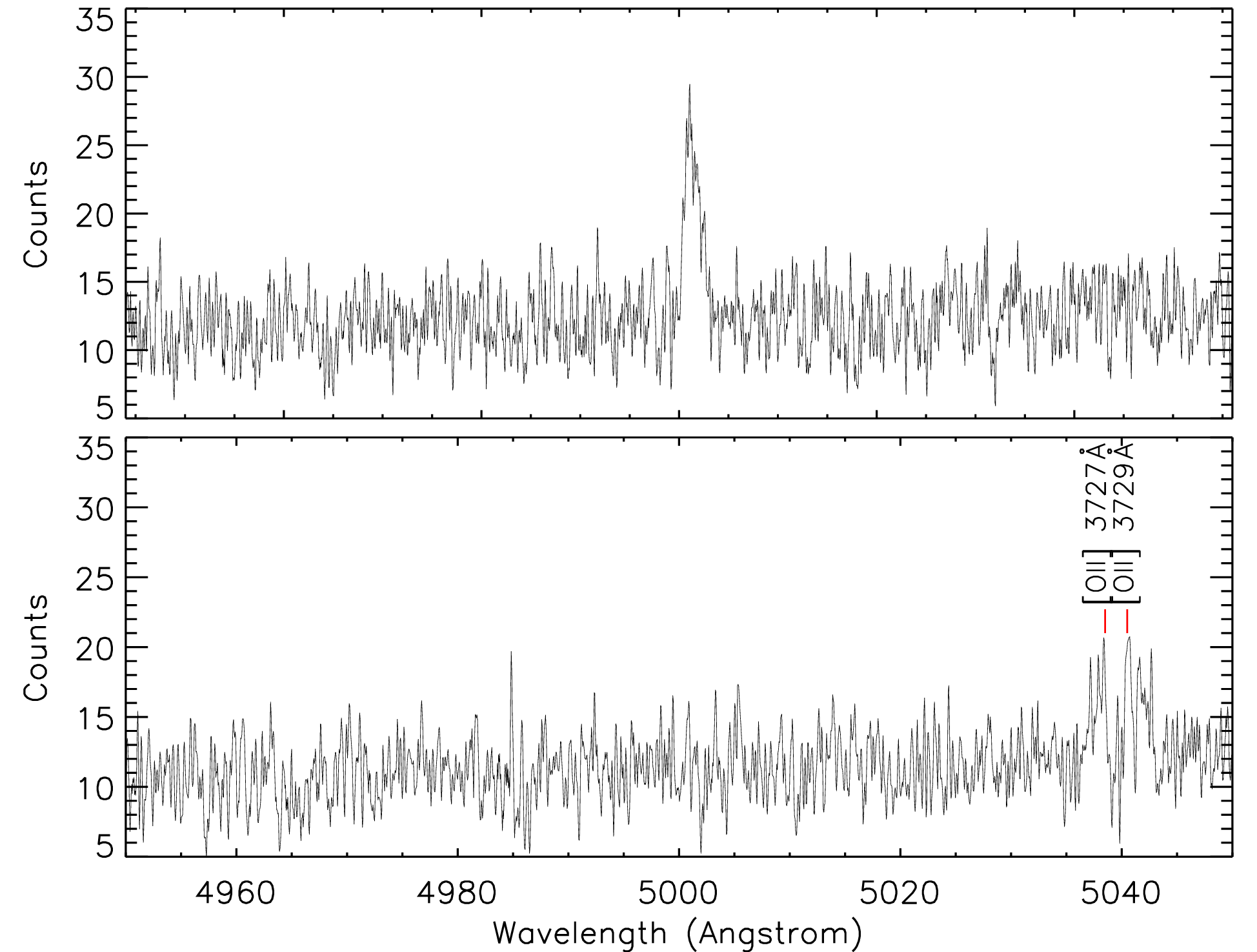}
   \caption{\small{Spectra for background emission line galaxies:
       Ly$\alpha$ emitter at $z \sim 3.1$ (top panel) and [OII]
       emitter at $z \sim0.34$ (bottom panel). Red vertical lines mark
       the positions of the two components of the $\lambda$3726+3729\AA\
       emission. We smoothed the spectra to 0.015 nm per pixel.}}
\label{spectra_example_EG}
 \end{figure}

The colour selection criteria are based on the strong [OIII]
$\lambda$5007\AA\ emission of a PN, with faint or no continuum.
Nonetheless, background galaxies like Ly$\alpha$ emitters at $z \sim
3.1$ and [OII] $\lambda$3727.26\AA\ emitters at $z \sim 0.34$ have
relatively strong lines that fall within the bandpass of the
narrowband filter. Thus, we classify the extracted spectra on the
basis of the shape of the line profile of the strongest emission. The
extracted spectra fall into these categories:
\begin{itemize}
\item[-] PN spectra: the [OIII]$\lambda$5007\AA\ emission of a PN is
  characterised by a narrow and symmetric line shape and very low
  continuum. In high S/N spectra, we detected the redshifted [OIII]
  $\lambda$4959/5007\AA\ doublet. Typical S/N ratios for the
  spectroscopically confirmed PN [OIII]]$\lambda$5007\AA\ cover a
  range of $2.5 \le \mathrm{S/N}\le 8.5$ per resolution element. In
  Fig.~\ref{spectra_example_PN} we show examples of single PN spectra
  with different V$_{\mathrm{LOS}}$.

  \item[-] Ly$\alpha$ spectra: the emission line of a Ly$\alpha$
    emitter has a broader and more asymmetric line profile,
    characterised by a steep drop-off at bluer wavelengths. This kind of
    signature comes from the forest absorption bluewards of
    Ly$\alpha$: the symmetric emission line is truncated below the
    object redshift by Ly$\alpha$ scattering in the intergalactic
    medium. Fig.~\ref{spectra_example_EG} shows an example of an
    extracted FLAMES spectrum for a Ly$\alpha$ emitter.

  \item[-] [OII] spectra: the [OII]$\lambda$3727\AA\ emitters are
    characterised by the redshifted, resolved, and broad emission lines
    of the oxygen doublet at $\lambda$3726-3729\AA. 
    Fig.~\ref{spectra_example_EG} shows an example of an
    extracted FLAMES spectrum for an [OII] line emitting galaxy.
\end{itemize}

 The final sample of emission line objects consists of 380 sources, of
 which 287 were classified as PNs and the remaining as background
 emission line galaxies, either as Ly$\alpha$ or as [OII]
 emitters. This is the largest sample of spectroscopically confirmed
 PNs around M87 thus far, which is about a factor 15 larger than the sample of
 \citet{doherty09}.

 The fraction of background emitters is consistent within one $\sigma$
 with the estimate in the photometric study of \citet{longobardi13}.
 They estimated that $\sim25\%$ of the total imaging sample
   would be background emission galaxies. Here we find that the same
   fraction of the automatic catalogue is in fact L$\alpha$
   or [OII] emitters.

 \subsection{Accuracy of the velocity measurements}\label{vacc}

 From the repeat observations of the same candidates in areas where
 different FLAMES plate configurations overlap (see
 Fig.~\ref{FLAMES_configuration}), we obtained independent velocity
 measurements for a subsample of PNs. The median deviation of these
 measurements is 4.2 kms$^{-1}$ and the whole distribution covers a
 range of $0.6 \le \Delta V_{\mathrm{LOS}}\le 16.2$ kms$^{-1}$. The
 largest errors occur when a cosmic ray falls near the wavelength of
 the [OIII]$\lambda$5007\AA\ emission in one of the exposures.

 \section{Halo and IC PN components}
\label{sec3}
When studying the outer regions of M87 out to a distance of $\sim 150$
kpc from the galaxy centre, we are tracing the light in the radial
range where the M87 halo blends into the ICL. \citet{arnaboldi04}
showed that the M87 stellar halo and the Virgo core ICL coexist for
distances $>60$ kpc from the galaxy centre, and \citet{doherty09}
showed that the two components overlap out to 150 kpc.
\citet{longobardi13} showed that the observed slope of the PN number
density profile is consistent with the superposition of two PN
populations associated with the M87 halo and ICL, respectively.

In the following subsections, we show that the distribution of LOS
velocities obtained from the FLAMES spectra shows evidence for two
dynamically distinct PN populations at large radii, confirming this
interpretation.

\subsection{Projected PN phase-space diagram}
\begin{figure}[h] \centering
   \includegraphics[width=9cm]{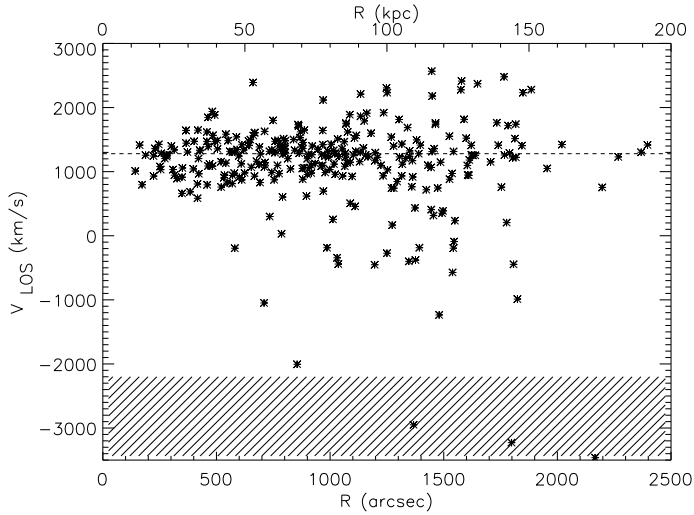}
   \caption{Projected phase-space diagram showing
     $\mathrm{V}_{\mathrm{LOS}}$ vs.~major axis distance $R$ from the
     centre of M87, for all spectroscopically confirmed PNs (black
     asterisks). The major axis distance is given both in
       arcsec (bottom axis), and in kpc (top axis), where $\rm{73\,
         pc=1\arcsec}$. The dotted horizontal line shows M87's
     systemic velocity V$_{\mathrm{sys}}=1275$ kms$^{-1}$ from
       Sect.\ref{sec3.2}. The shaded area represents the region of
     the projected phase-space where the blue-shifted [OIII] 4959\AA\
     emission line would fall below the wavelength of the blue edge of
     the FLAMES sort ordering filter HR08.}\label{Pspace_data}
\end{figure}

For each confirmed PN spectrum, we measured
$\mathrm{V}_{\mathrm{LOS}}$ and computed the major axis distance via
the formula $R^2=\mathrm{x_{PN}{^2}/(1-e)^{2}+y_{PN}^{2}}$, where
$\mathrm{e}$ is the isophote's ellipticity from \citet{kormendy09} and
$\mathrm{x_{PN}, y_{PN}}$ are the PN coordinates measured in a
reference frame centred on M87, where the y axis is aligned with the
major axis of the outer elliptical isophotes at $\rm{P.A.=-25.6}$
\citep{kormendy09}. In Fig.~\ref{Pspace_data}, we show the projected
phase-space diagram V$_{\mathrm{LOS,PN}}$ vs. $R$ for the
spectroscopically confirmed PNs in the M87 survey (black asterisks).

\begin{figure}\centering
   \includegraphics[width=9.cm]{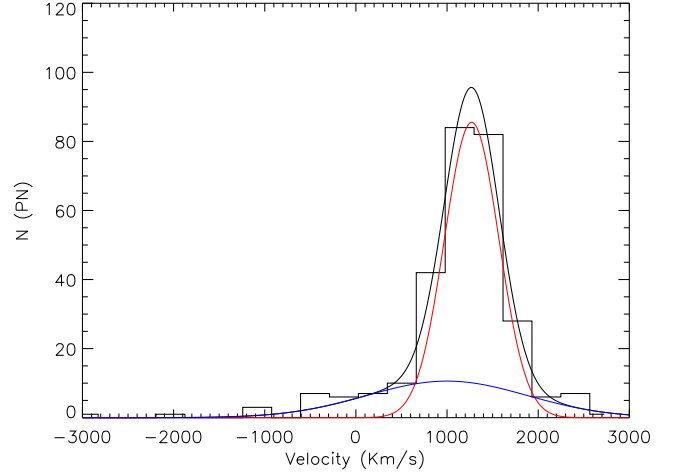}
   \caption{\small{Histogram of the line-of-sight velocities of the
       spectroscopically confirmed PNs (black histogram) fitted with a
       double Gaussian (black curve). Red and blue lines represent the
       two Gaussians associated with the M87 halo and the IC
       components.}}\label{V_hist}
\end{figure}

In this projected phase-space, PN velocities show a concentration
around the systemic velocity of M87 (V$_{\mathrm{sys}}=1275$
kms$^{-1}$; see Sect.~\ref{sec3.2}), in addition to a scattered
distribution at higher and lower velocities.  In Fig.~\ref{V_hist}, we
show the histogram of the velocities for the entire sample. It
  has two very strong and quite asymmetric wings around the main
  peak. The wing (or tail) at low velocities is more extended than
  the tail at high velocities with respect to the peak.  These
  extended tails are very different from those measured in the LOS
  velocity distributions (LOSVDs) of isolated early-type galaxies,
  which are well described by single Gaussian distributions to within
  $\sim 1\%$ \citep[e.g.,][]{gerhard93,bender94}.  The total velocity
  distribution is well fitted by a double Gaussian, while a single
  Gaussian is a poor fit to the observed LOSVD. The reduced $\chi^2$
  of a double Gaussian fit is 1.1, while for a single Gaussian it is 2.0.
  Therefore, we fit the LOSVD in Fig. 5 with the sum of two
  Gaussians. The fit to the data returns a narrow component centred
on V$_{\mathrm{LOS,n}}= 1270.4$ kms$^{-1}$ with velocity dispersion of
$\sigma_{\mathrm{n}}=298.4$ kms$^{-1}$, and a broad component, centred
on V$_{\mathrm{LOS,b}}= 999.5$ kms$^{-1}$ with a larger velocity
dispersion $\sigma_{\mathrm{b}}=881.0$ kms$^{-1}$.

The broad component is shifted from the M87 systemic LOS velocity:
both V$_{\mathrm{LOS,b}}$ and $\sigma_{\mathrm{b}}$ are consistent
with those values determined for the LOSVD of galaxies in the main
sub-cluster region $A$ of the Virgo cluster \citep{binggeli93,conselice01}.  The
LOSVD around M87 is thus bimodal, containing a narrow component
associated with the systemic velocity of the galaxy plus a broader
component associated with the ICL.
 
\label{sect_3_1}

Different LOSVDs for the halo and ICL are predicted by cosmological
analysis of structure formation. Using hydrodynamical
  cosmological simulations, \citet{dolag10,cui14} study the LOSVD of
  star particles at the centre of their clusters. These authors find
  that this LOSVD is well described by the sum of two Gaussians, with
  similar average velocities but different $\sigma$. One component is
gravitationally bound to the galaxy and more spatially concentrated;
the other is more diffuse and its high-velocity dispersion reflects
the satellites' orbital distribution in the cluster gravitational
potential.  It is plausible that the halo stars are spatially confined
as a consequence of merging processes that led to the formation of the
BCG \citep{murante07,contini14}.

\subsection{Robust separation of the M87 halo and ICL}\label{sec3.2}

The overall LOSVD of the PNs in M87 is characterised
by a narrow component associated with the M87 halo, superposed on a
broad IC component with a shifted mean velocity and much larger
velocity width. We are now interested in separating halo and IC PNs
based on their different LOSVDs. In this analysis, we combine our PN
sample with that of \citet{doherty09}. We concentrate on those PNs
that have a major axis distance $R \le 190$ kpc to study the
transition between the M87 halo and the Virgo ICL. The combined total
sample consists of 299 PNs with measured positions and velocities
within $R \le 190$ kpc; the PNs from \citet{doherty09} further out are
classified as ICPNs and are not discussed further.

\begin{figure}[h] \centering
   \includegraphics[width=9cm]{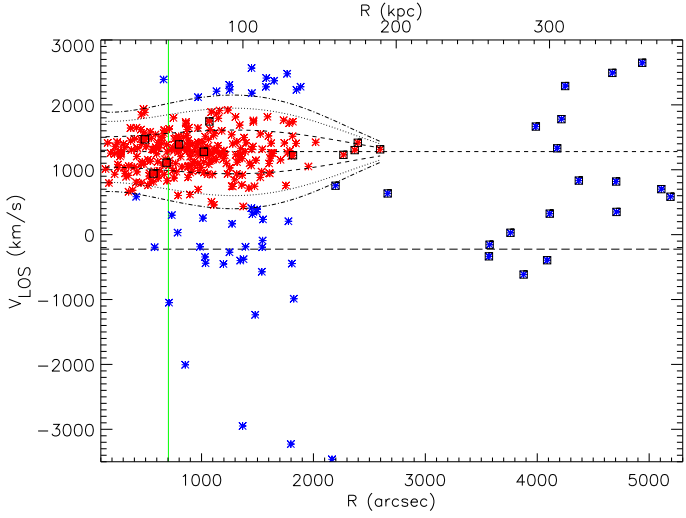}
   \caption{Projected phase-space diagram,
     $\mathrm{V}_{\mathrm{LOS}}$ vs.\ major axis distance from the
     centre of M87, for all spectroscopically confirmed PNs from this
     work and \citet{doherty09}.The major axis distance is
       given both in arcsec (bottom axis), and in kpc (top axis),
       where $\rm{73\, pc=1\arcsec}$. The PNs are classified as M87
     halo PNs (red asterisks) and ICPNs (blue asterisks),
     respectively; see text. Black squares identify spectroscopically
     confirmed PNs from \citet{doherty09}. The smoothed 1, 2, and 2.5
     $\sigma$ thresholds are represented by the dashed, dotted, and
     dot-dashed lines, respectively. The dashed horizontal line shows
     the M87 systemic velocity V$_{\mathrm{sys}}=1275$ kms$^{-1}$ as
     computed in Section~\ref{sec3.2}, while the continuous
       green line shows the effective radius $R_{e}=703.914 \arcsec$
       determined by \citet{kormendy09}.  At V$_{\mathrm{LOS}}=
     -220\, {\rm kms}^{-1}$, we plot the M86 systemic velocity (long
     dashed line).}\label{Pspace}
 \end{figure}

To separate halo and IC components, we use a sigma clipping
algorithm in elliptical radial bins, corresponding to vertical strips
in the projected phase-space diagram in Fig.~\ref{Pspace}.  The idea
is to separate the velocities in the narrower Gaussian (see
Fig.~\ref{V_hist}) from those in the high- and low-velocity wings of
the distribution. By tagging our sources depending on whether their
$\mathrm{V_{LOS}}$ belongs to the narrower or wider Gaussian, we
assign PNs to either the M87 halo or the ICL.

\emph{The velocity dispersion profile of the M87 halo: a robust sigma
  estimate--} We binned the PN velocity sample in elliptical annuli
and, for each bin, we determined the standard deviation of the LOSVD
for the PNs in this bin. We applied a $2\sigma$ limit with respect to the systemic
velocity of M87\footnote{
The systemic velocity of M87 is taken as the median
  value of the entire sample of velocities within two sigma of the
  original median. Our value $\rm{1275 \pm 24\, kms^{-1}}$  is
    consistent with those obtained by \citet{binggeli87}
    ($\rm{V_{sys}=1258 \pm 10 kms^{-1}}$) , and \citet{cappellari11}
    ($\rm{V_{sys}=1284 \pm 5 kms^{-1}}$). }, V$_{\mathrm{sys}}=1275 \pm 24$ kms$^{-1}$, and calculated the
  dispersion for all PNs with $|$V$_{\rm LOS}$-V$_{\rm
    sys}| <2\sigma\, {\rm kms}^{-1}$. We then scaled this dispersion
by a numerical factor determined from Monte Carlo simulations to
correct to the dispersion of a complete Gaussian distribution~\citep[see][]{mcneil10}.
Since we expect the initial estimate for the
$2\sigma$ to be influenced by the ICPNs, we repeat this process until
the dispersion value stabilises.

\emph{Separating M87 halo and IC PNs: sigma clipping --} We now
identify the two components using a sigma clipping algorithm.  We
begin by classifying as ICPNs all velocity outliers that deviate from
the M87 systemic velocity V$_{\mathrm{sys}}=1275$ kms$^{-1}$ by more
than 2$\sigma$. To obtain the required $\sigma$ value at the radius of
each PN, we took the robust estimates of the velocity dispersions in
the elliptical radial bins, and fitted these data with a fourth-order
polynomial. This takes radial gradients into account and, at the same
time, reduces the effects of binning and scatter in the dispersion
profile on the separation of the components. Using the $2\sigma$
threshold from the interpolated polynomial, we identify 243 objects as
M87 halo PNs.

Two further steps are still needed. Firstly, the 2$\sigma$ criterion accounts
for $\sim 95.5$\% of a complete Gaussian distribution; hence we expect
it to have missed 11 halo PN candidates.  To include these, we
considered all outliers within $3\sigma$ from the M87
$\mathrm{V_{sys}}$ and, from those, selected the 11 with the smallest
$|$V$_{\rm LOS}$-V$_{\rm sys}|/\sigma$ ratios. This leads to a final
sample of 254 M87 halo PNs and 45 ICPNs. In this final sample, 11 halo
PNs and 1 ICPN came from \citet{doherty09}.

Secondly, as can be seen in Fig.~\ref{V_hist} the ICL and halo
velocity distributions overlap, and as result, when using the
sigma-clipping algorithm, the ICPNs at low velocities relative to
V$_{\mathrm{sys}}\sim1275$ kms$^{-1}$ are considered to be part of
the halo component. To statistically quantify this effect, we compared
the halo and IC velocity distributions in each radial bin (or slice in
the phase-space). We approximated the halo distribution in the bin as
a Gaussian centred on the systemic velocity of M87 with a dispersion
equal to the average value for that bin, and for the IC component we
used the same Gaussian in all radial bins using the same parameters as in
Fig.~\ref{V_hist}. We then calculated the fraction of ICPNs that lie
inside the halo distribution as the area of overlap between the two
curves. With this analysis we obtain a statistical estimate of the
number of ICPNs contained in the halo velocity distribution in each
bin. For all bins combined, we find that a further $\sim$17\% of the
halo sample, i.e., $\sim 44$ PNs, are to be associated with the IC
component.

In Fig.~\ref{Pspace}, we show the projected phase-space distribution
of our PNs as in Fig.\ref{Pspace_data}, with velocities colour-coded
to show the membership to the halo (red) and ICL (blue). Since we
only know statistically, but not individually, which PNs in the halo
velocity range are ICPNs, these are also shown with the red halo
colour.

In Fig.~\ref{Pspace} we also show 1/2/2.5$\sigma$ limit contours for
the halo PNs, obtained by fitting a polynomial to the $\sigma$ values
from the robust estimation. The velocity dispersion profile increases
from 250$\arcsec$ to 1200 $\arcsec$, and then decreases, showing a
colder component at radii $ R > 1200 \arcsec$. A more
detailed analysis of the dispersion profile will be provided in a
separate paper (Longobardi et al., in preparation).

We now enquire whether any ICPNs could be associated with other
galaxies in the cluster.  The first to consider is M86, because it is
bright, has the most extended halo, and is relatively close to M87
($\sim 1.2^{\circ}$ away).  Extrapolating the Sersic fit from
\citet{janowiecki10}, we compute the total luminosity associated with
the M86 halo in our surveyed area to be $\rm{L_{V,bol}\sim 8.0 \times
  10^{8} L_{\odot,bol}}$.  We adopt a similar PN specific number as
for M87, $\rm{\alpha_{2.5}=1.01\times 10^{-8} PN/L_{\odot,bol}}$ (see
Sect.\ref{sec5.1}), but we correct for the fact that we can only sample
the brightest 2.3 mag of the PNLF because of the larger distance to
M86 \citep{mei07}.  This leads to a predicted number of 7 M86 PNs that
could be counted as IC component, i.e., only $\sim2.5\%$ of the
completeness-corrected ICPN sample (see Sect.\ref{sec5.1}), or 2 out
of a total of 88 ICPNs.

Other fainter galaxies have less extended halos and contribute less to
the sampled ICPNs around M87 unless they are very near or in the
survey area. Moreover, from the shape of the ICPNLF (see
Sect.\ref{sec5.2}) the majority of ICPNs is likely to have been
stripped from low-luminosity star-forming galaxies, such as M33 and the
LMC.  Any ICPNs still bound to dwarf galaxies would be correlated with
their position and velocity. Only three PNs with velocity V$\sim$350 km/s
out of the ICPN sample seem correlated with the position and systemic
velocity of a dwarf galaxy, IC3549, which has $\rm{V_{LOS}=375\,
  kms^{-1}}$ \citep[][]{mcCarthy09}. Hence, we can state that the
contribution from other nearby galaxies to the ICPN sample is
negligibile and does not affect any of the results we present.

Finally, a few ICPNs are characterised by
extraordinary blue shifts relative to M87 ($\mathrm{V_{LOS}<-1000}\,
{\rm kms}^{-1}$). These (hyper-) velocities of $|$V$_{\rm LOS}$-V$_{\rm
  sys}| >2300\, {\rm kms}^{-1}$ relative to M87, corresponding to
several $\sigma_{\rm Virgo}$, could be because of infall from the
outskirts of the Virgo cluster, perhaps associated with the infall of
the M86 group, or they could be tracers of three-body interactions, as
previously hypothesised for the extreme globular cluster observed at a
projected distance of $\sim 80$ kpc from M87 with a velocity of
$\mathrm{V_{LOS}<-1025}$ kms$^{-1}$ \citep{caldwell14}.

\subsection{Spectroscopic validation of the PN subsample}\label{sec3.3}

In Section~\ref{section2.4} we classified 287 spectra as PNs on the
basis of the line profile of the strongest emission.  We now
strengthen this earlier classification based on the detection of the
weaker 4959\AA\ line of the [OIII] doublet in the PN spectra.

In 114 out of the 287 spectra, nearly 40\% of the sample, we are able
to detect the Doppler-shifted [OIII]$\lambda$4959\AA\ line with the
expected \rm{1:3} ratio of the [OIII]$\lambda$4959/5007\AA\ line
fluxes. In the rest of the sample, the single spectra do not have the
required S/N ($\rm{S/N > 3}$) for the main [OIII]$\lambda$5007
  \AA\ emission line  to allow the detection of the weaker 4959\AA\
line.  However, following \citet{arnaboldi03} we can achieve the
required S/N by stacking these spectra, after shifting their
[OIII]$\lambda$5007\AA\ emission to a common reference wavelength.  By
measuring the [OIII]$\lambda$5007/4959\AA\ line ratio of the coadded
spectrum, we can statistically constrain the fraction of misclassified
PN spectra: if the stacked spectrum contains misclassified PN
candidates, the ratio [OIII] $\lambda$5007/4959\AA\ is
larger than three.

\begin{figure}\centering
   \includegraphics[width=7.5cm, clip=true, trim=0cm 2cm 0cm
   0cm]{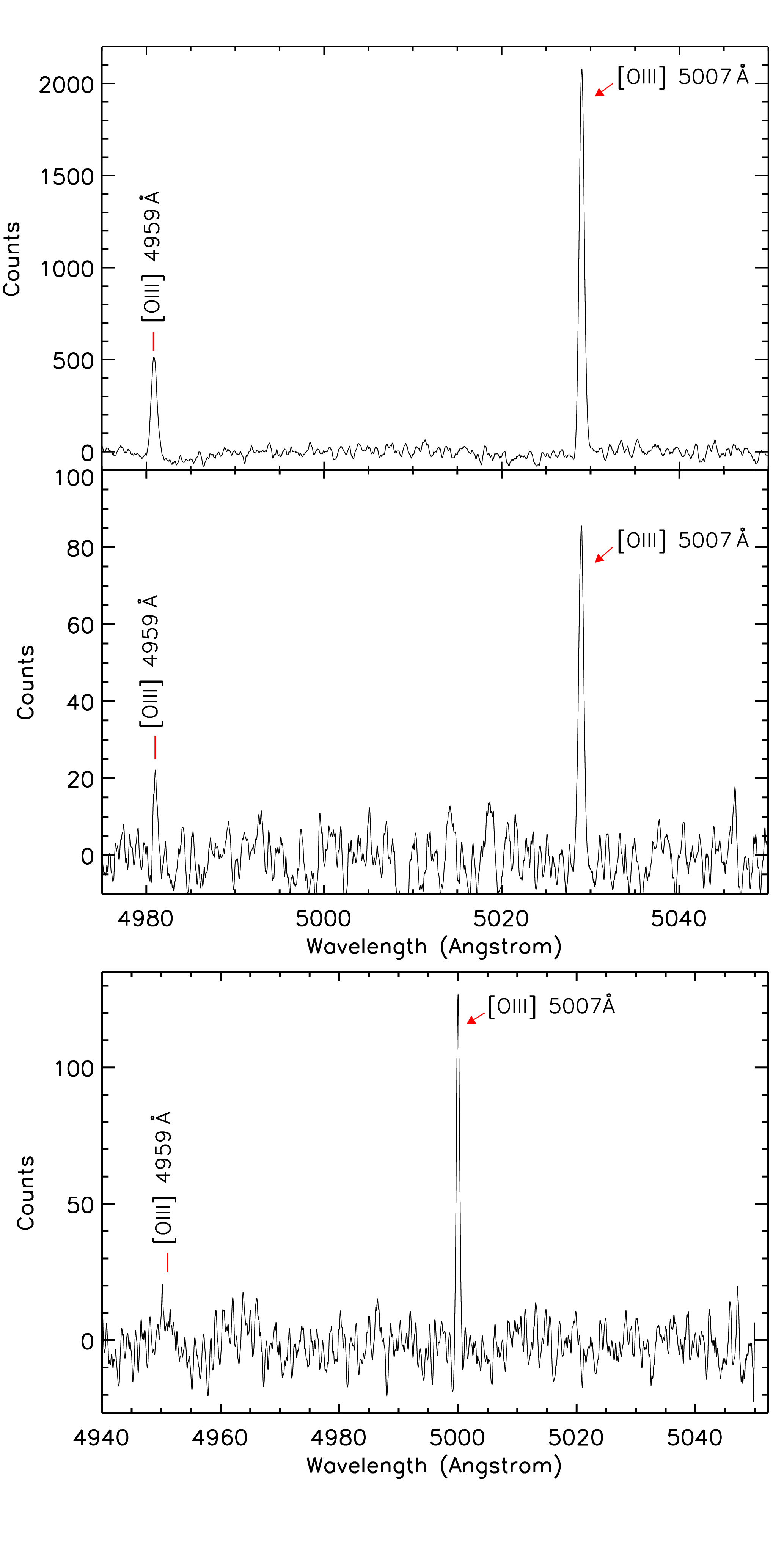}
   \caption{\small{Combined spectra for the spectroscopically
       confirmed sub-samples of PNs belonging to the halo (upper
       panel), IC high-velocity (central panel) and IC low-velocity
       (bottom panel) classes. As expected from atomic physics, the
       [OIII]$\lambda$4959/5007\AA\ doublet is
       detected with a flux ratio 1:3 confirming that the large
       majority of our candidates are true PNs (see text for more
       details).}}\label{spectra_comb}
 \end{figure}

The 287 spectra are further grouped in three classes: M87 halo
spectra, IC high-velocity, and IC low-velocity spectra. We have 
254, 13, and 31 PN in each category, respectively. Three sources in the
IC low-velocity class are not included because their Doppler shifted
[OIII]$\lambda$4959\AA\ emission fall at a shorter wavelength
than the blue edge of the FLAMES HR08 filter (see
Fig.~\ref{Pspace_data}).  The spectra are shifted so that the main
[OIII]$\lambda$5007\AA\ emission falls at the same nominal wavelength
for all spectra of a given class.  For the halo and IC high-velocity
class, we adopt the nominal wavelength of 5029\AA\ for
[OIII]$\lambda$5007\AA, i.e. the redshifted wavelength at the systemic
velocity of M87. For the IC low-velocity class, the adopted nominal
wavelength is 5000\AA. We shifted the single spectra for each class,
and obtained three coadded spectra. These are plotted in
Fig.~\ref{spectra_comb}. The [OIII] doublet is visible in all spectra,
with the [OIII]$\lambda$4959\AA\ line visible at the correspondent
shifted wavelength.  In all three spectra, the FWHM of
[OIII]$\lambda$5007\AA\ is FWHM=0.6\AA\ (Table~\ref{obj_frac}),
somewhat smaller than the typical FWHM of single spectra, FWHM=0.8
\AA.  Because of filter edge effects, together with the
shifting and alignment of the spectra with reference to the stronger
line, the weaker [OIII]$\lambda$4959\AA\ emission is broader in the IC
low-velocity spectrum than in the single PN spectra.

The fluxes and FWHMs of the [OIII]$\lambda$4959/5007\AA\
emissions in the coadded spectra are measured via a Gaussian fit to
the lines with the IRAF task \textit{splot}. The errors on the line
fluxes were calculated using the relation \citep{perez03}
\begin{equation}
\sigma_1=\sigma_{c}N^{1/2}[1+EW/(N\Delta)]^{1/2},
\label{flux_err}
\end{equation}
where $\sigma_{1}$ is the error in the line flux, $\sigma_c$ is the
standard deviation in a box near the measured line and represents the
error in the definition of the continuum, $N$ is the number of pixels
used to measure the line flux, $EW$ is the equivalent width of the
line, and $\Delta$ is the wavelength dispersion in \AA\
pixel$^{-1}$. In Table~\ref{obj_frac} we give a summary of the observed
properties of the [OIII] doublet of the coadded spectra.

The [OIII] doublet is visible in the three spectra from the different
classes, with approximate flux ratio of 3:1, as expected from
  atomic physics.  Given the uncertainties on the measured fluxes,
the spectral validation shows that at $1\sigma$ $>99\%$ of the halo
and $>90\%$ of the IC high-velocity spectra are true PNs. This
  implies that only two halo PNs and one high-velocity ICPN
  could be misclassified. For the IC low-velocity PNs, the line ratio
  of the [OIII]5007/4959 \AA\ is $3.1\pm 0.3$ from
  Table~\ref{obj_frac}. Hence up to 13\% of the spectra could be
  misclassified, i.e. four ICPNs. On the total sample of 287
spectroscopically confirmed PNs, at most seven candidates might have been
misclassified, i.e. only 2\% of the whole sample.

  \begin{table*}
   
    % \centering
  % \caption{\small{Flames Configuration and Exposure Time}}
   \centering

   \caption{\small{Fluxes, FWHMs, and line ratios of the [OIII] $\lambda$5007/4959\AA\ 
     doublet for halo, high-velocity and low-velocity IC PNs sub-samples.}}
  \label{obj_frac}
     \begin{tabular}{c c c c c c}
      
     \hline
     \hline\\

     IDs PN sub-sample &Line Flux$_{5007}$& Line Flux$_{4959}$&  FWHM$_{5007}$ & FWHM$_{4959}$& Line Ratio \\
     &( Counts )& ( Counts ) & ( \AA\ )& ( \AA\ )& \\
     &(1)& (2) & (3)& (4)& (5)\\
 
          \hline\\

      halo & 1528.0$\pm$0.2& 506.9$\pm$0.3&0.60$\pm$0.08&0.76$\pm$0.1&3.01$\pm$0.02 \\
      IC high-velocity   &  64.1$\pm$3.0&  20.6$\pm$1.2&0.60$\pm$0.07&0.6$\pm$0.09&3.1$\pm$0.2\\
      IC low-velocity &   106.3$\pm$3.0&  33.9$\pm$3.7&0.60$\pm$0.08& 1.4$\pm$0.1&3.1$\pm$0.3\\
    
     \hline\\

   \end{tabular}
   \tablefoot{Columns (1) and (2): fluxes for the [OIII] main and second emission line, respectively. 
     Errors are calculated following Eq.\ref{flux_err}.\\
     Columns (3) and (4): emission line FWHMs from Gaussian fitting of
     the spectral lines. Errors represent 1$\sigma$ uncertainties from
     sampling statistics.\\
     Column(5): flux ratio between the [OIII] doublet emission lines. Errors are propagated from the flux errors.}
 \end{table*}

 \section{Halo and IC density profiles} 
\label{sec4}

By tagging PNs according to their V$_{\mathrm{LOS}}$, we differentiated 
between halo and IC PNs in the last section. We are now interested in
recovering the spatial distribution of these two components and in
studying their number density profiles separately. 

\begin{figure}
\centering 
\includegraphics[width=8cm]{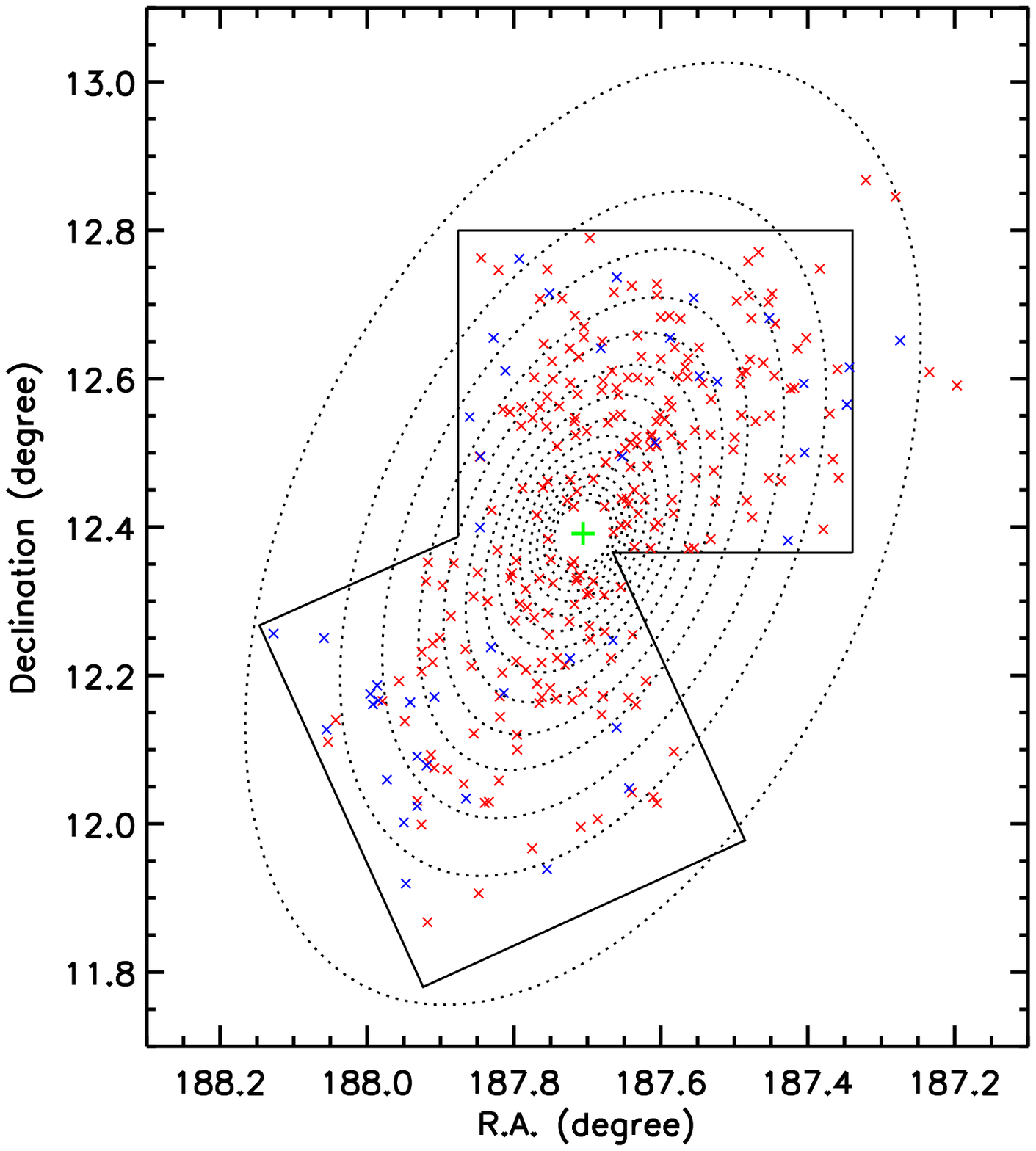} 
\caption{Sky positions of the spectroscopically confirmed halo (red
  crosses) and IC PNs (blue crosses).  The green plus sign indicates
  the centre of M87.  The ellipses (dotted lines) trace the M87
  isophotes between photometric major axis radii $R=2.'8$ and
  $R=40.'7$, for position angle P.A.=-25.6$^{\circ}$, from
  \cite{kormendy09}.  The solid squares depict the area covered by our
  narrowband imaging survey \citep{longobardi13}. North is up, East
  to the left.}
   \label{spatial_dist}
 \end{figure}
   
 In Fig.~\ref{spatial_dist} we show the sky positions of the
 spectroscopically confirmed halo and IC PNs.  If the halo and ICL are
 characterised by different evolutionary histories, their radial
 distributions may be different. If so, this can be seen in the PN
 density profile. Because PNs follow light, the presence of a single
 parent stellar population would be reflected in a PN number density
 distribution proportional to the surface brightness profile at each
 radius.  Deviations of the PN number density from the light profile
 would trace the presence of more than one stellar population in
 the surveyed region \citep{longobardi13}.

\subsection{Completeness corrections}

To construct a PN density profile, we require a spatially complete
sample hence we must correct the detected PN number using a
completeness function, C$_{\mathrm{tot}}(\mathrm{x_{PN},y_{PN}})$,
which accounts for the selection function of the sampled PNs over the
surveyed area. Our PN sample is affected by four different kinds of
incompleteness, which are related to the photometric identification of
the candidates and the selection effects in the spectroscopic
observations. These are:
\begin{itemize}
\item[]\textbf{Photometric incompleteness}, characterised by:
\begin{itemize}
\item[i)] C$_{\mathrm{phot,sp}}(R)$ - Spatial incompleteness due to
  the high galaxy background and foreground stars that affect the
  detection of PN candidates in the images. This was estimated in
  \citet{longobardi13} by adding a simulated PN population to
  the scientific images and determining the fraction of simulated
  objects recovered by SExtractor in the different elliptical annuli
  shown in Fig.~\ref{spatial_dist}.
\item[ii)] C$_{\mathrm{phot,col}}$ - Colour incompleteness due to the
  colour criteria adopted for the automatic selection of PN
  candidates. This colour incompleteness was computed by analysing the
  properties of the recovered simulated PN population in the
  colour-magnitude diagram \citep[see][for more detail]{longobardi13}. 
  This incompleteness affects only the M87SUB FCEN and M87SUB FEDGE
  fields, where the photometric candidates were selected through
  colour criteria (see Sect.~\ref{sec2}), and is
  magnitude-dependent. For constructing the density profile, however, we use
  an average value, computed for the whole sample down to 2.5
  magnitudes below the bright cut-off, which amounts to $0.7$.
\end{itemize} 
\vspace{0.2 cm}
\item[]\textbf{Spectroscopic incompleteness}, characterised by:
\begin{itemize}
\item[iii)] C$_{\mathrm{spec,sp}}(R)$ - Spatial incompleteness due to
  the limited number of fibres (up to 132) that can be allocated
  for each FLAMES field. This was estimated by computing the ratio
  between the number of allocated fibres and the total number of
  photometric candidates in each elliptical annulus shown in
    Fig.~\ref{spatial_dist}.
\item[iv)] C$_{\mathrm{spec,fb}}(\mathrm{x_{PN},y_{PN}})$ -
  Incompleteness due to fiber-target misalignment. This incompleteness
  was estimated from the detection statistics of objects in common
  between overlapping FLAMES plate configurations, whose [OIII]
  emission was detected in spectra taken with either one or two
  of the two-plate configurations (see Sect.~\ref{section2.3}).
\end{itemize}
\item[] Our spectra are sufficiently deep that we found no dependence
of the spectroscopic incompleteness on the magnitude of the PN candidates.
\end{itemize} 

The total completeness function,
C$_{\mathrm{tot}}(\mathrm{x_{PN},y_{PN}})$, is the product of the
photometric and spectroscopic incompleteness:
\begin{eqnarray}
  C_{\mathrm{tot}}(\mathrm{x_{PN},y_{PN}})&=& \mathrm{C}_{\mathrm{phot,sp}}(R)*\mathrm{C}_{\mathrm{phot,col}}     \nonumber \\
   &*& \mathrm{C}_{\mathrm{spec,sp}}(R) * \mathrm{C}_{\mathrm{spec,fb}}(\mathrm{x_{PN},y_{PN}}),
\end{eqnarray}
with
$$
\mathrm{C}_{\mathrm{phot,col}} = \left\{ \begin{array}{ll}
 \mathrm{C}_{\mathrm{phot,col}} &\mbox{ for M87SUB FCEN, M87SUB FEDGE}, \\
  1 &\mbox{ elsewhere}.
       \end{array} \right.
$$ 
The completeness-corrected number of PNs in each bin of major axis
distance $R$ is then
\begin{equation}
\mathrm{N_{c}(R)=}\Sigma_{i}^{N_{\mathrm{obs}}} k_{i}(R),
\label{eq:Nc}
\end{equation}
where the sum extends over all PNs of the halo or IC component in the bin,
respectively, and $k_{i}(R)=1/{\rm C}_{\mathrm{tot}}(\mathrm{x_{PN_{i}},y_{PN_{i}}})$ 
is the completeness-corrected specific weight of each observed PN at
its position.

\subsection{Density profiles of halo and IC component}

To construct the PN number density profile and compare it to
the galaxy's surface brightness profile, we bin our PN sample in
elliptical annuli. The radial range of the elliptical annuli is chosen
such that they include the major axis distance of the innermost and
outermost PN candidates in the photometric sample of
\cite{longobardi13}. Their P.A.s and ellipticities are taken from
\cite{kormendy09}. The sizes of the annuli are determined separately
for the halo and IC components, such that for each component all bins
contain at least ten spectroscopically confirmed PNs.  In each annulus
and for each component, we compute the completeness-corrected PN
number density as the ratio of the completeness-corrected number of PN
(Eq.~\ref{eq:Nc}) and the area of the portion of the annulus
intersecting our FOV\footnote{These areas are estimated using Monte
  Carlo integration.},
\begin{equation}\label{NPNR}
\sigma_{\mathrm{PN}_{j}}(R)=\frac{\mathrm{N}_{\mathrm{c},j}(R)}{A(R)}.
\end{equation}
Here we consider all the spectroscopically confirmed PNs whose
magnitudes are within 2.5 mag below the bright cut-off. The subscript
$j$ indicates the two different PN components, halo and ICL.  The
major axis distances $R$ in Eq.~\ref{NPNR} are determined by computing
the average major axis distance of all PNs falling within each
elliptical annulus.

\noindent
\begin{figure}
  \centering
  \includegraphics[width=8.5cm]{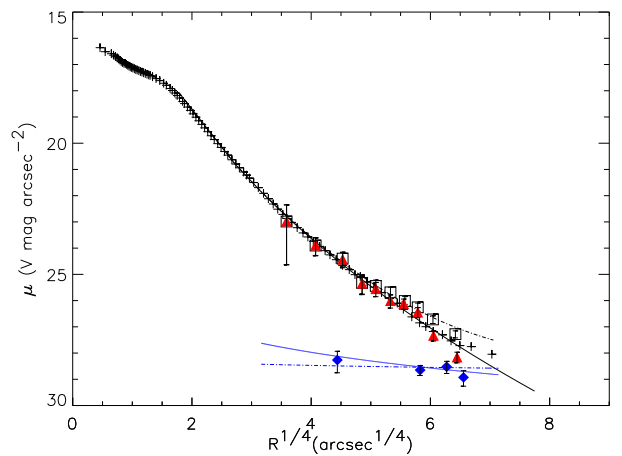}
  \caption{Logarithmic PN number density profiles for the M87 halo
    (red triangles) and the IC PNs (blue diamonds), corrected for
    incompleteness. The error bars include counting statistics
      and completeness correction.  The halo PN profile compares well
    with the surface brightness profile from \citet[][plus
    signs]{kormendy09}.  The continuous black line shows their Sersic
    fit with $n=11.8$.  The ICPN number density decreases towards
    larger radii as I$_{\mathrm{ICL}}\propto R^{\gamma}$ with $\gamma$
    in the range $[-0.34;-0.04]$ (full and dashed-dotted blue lines).
    Black squares show the combined halo and IC PN number density
    profile, which is well modelled by the two-component photometric
    model (dot-dashed black line, see Section \ref{sec5}).}
  \label{density_profile}
\end{figure}

In Fig.~\ref{density_profile} we show the comparison between the V-band
M87 surface brightness profile $\mu_{\mathrm{K09}}$ from
\citet{kormendy09}, with the logarithmic PN density profiles for the
halo (red triangles) and IC (blue diamonds) PNs, defined as
\begin{equation}
 \mu_{\mathrm{PN_{j}}}(R)=-2.5\log_{10}\left(\sigma_{\mathrm{PN_{j}}}(R)\right)+\mu_0.
  \label{rho}
\end{equation}
The value $\mu_0$ is a constant to be added so that the PN number
density profile matches the $\mu_{\mathrm{K09}}$ surface brightness
profile.  As described in Sect.~\ref{sec3.2}, the kinematic
decomposition of the halo and IC components does not identify ICPNs
in the velocity range of the halo, and their contribution must be
evaluated statistically for each bin. Averaged over the bins,
$\sim$10\% of the halo PN sample is thus estimated to be associated
with the ICL, in addition to the ICPNs identified from their large
velocities.  For each of the radial bins, we subtract the estimated
contribution from the halo component and add it to the IC component.
In Fig.~\ref{density_profile}, the PN number densities for halo and
ICL account for this effect. Furthermore, the profiles shown in
Fig.~\ref{density_profile} are computed using all PNs whose magnitudes
are within 2.5 magnitude from the bright cut-off. However, the halo and
IC number density profiles do not change significantly when the
fainter spectroscopically confirmed PN are also included.

We find that the $\mu_{\mathrm{PN_{halo}}}$ agrees well with the
surface photometry: the halo PN logarithmic number density profile
follows the surface brightness profile. The ICPN logarithmic number
density, as the halo profile, is centrally concentrated towards M87.
However, it has a flatter profile that decreases towards larger radii as
$I_{\mathrm{ICL}}\propto\mathrm{R}^{\gamma}$ with $\gamma$ in the
range $[-0.34;-0.04]$, depending on the choice of binning.  These
results are consistent with predictions from hydro-dynamical
simulations, where the radial density profile of the bound component,
i.e.\ the halo, is observed to be much steeper than that of the
diffuse IC component \citep{murante04,dolag10}.

\begin{figure}
  \centering
  \includegraphics[width=8.5cm]{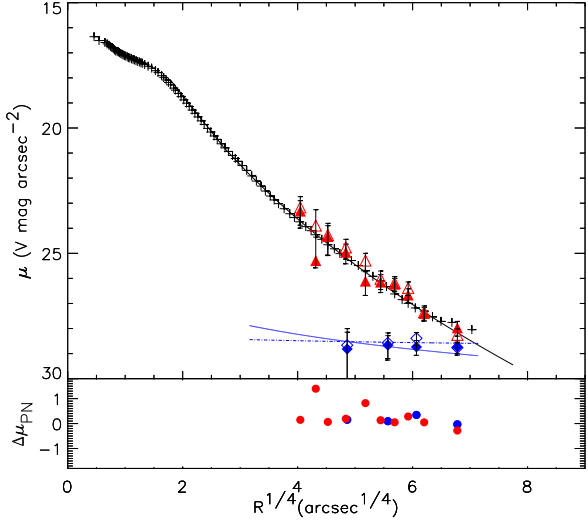}
  \caption{\textit{Top panel}: Same as in Fig.~\ref{density_profile}, but here
    the halo and IC number density profiles are computed separately
    for the NW (filled symbols) and SE (open symbols) sides of M87.
    \textit{Bottom panel}: Difference between the logarithmic PN number density
    profiles of the halo (red dots) and IC (blue dots) PNs in the NW
    and SE M87 regions. A value of zero indicates no difference in the
    number of sources on opposite sides at a given major axis
    distance.}
  \label{density_profile_NS}
\end{figure}

Finally, we compute density profiles for the PNs for the NW and SE
sides of M87 independently, to search for a NW-SE asymmetry in the
spatial distribution. For this test, we now use the total sample of PNs,
including very faint PNs, to increase the statistics in each
half annulus (this also changes the binning).
Fig.~\ref{density_profile_NS} shows the four number density profiles;
each pair of profiles are consistent with each other and the halo
profiles are consistent with the galaxy surface brightness
profile. Thus within the statistical uncertainties, the stellar halo
density is NW-SE symmetric and the PN number density follows the light
on both sides. The ICPN density profile pair are flatter than
the halo again, and consistent with the previous surface brightness profile
I$_{\mathrm{ICL}}\propto R^{\gamma}$ with $\gamma=[-0.34;-0.04]$; no
asymmetry is evident. In Fig.~\ref{density_profile_NS} we also plot
the differences between the NW and SE logarithmic densities as
function of radius (bottom panel). For both the halo (red filled
circles) and the ICL (blue filled circles), these differences are
consistent with zero within the uncertainties (see further discussion
in Sect.~\ref{sec6}).

 \section{Halo and IC populations}
\label{sec5}
\subsection{The $\alpha$ parameter}\label{alpha}
\label{sec5.1}
The total number of PNs, N$_{\mathrm{PN}}$, is proportional to the
total bolometric luminosity of the parent stellar population through
the luminosity-specific PN density, or $\alpha$-parameter for short,
such that N$_{\mathrm{PN}}=\alpha L_{\mathrm{bol}}$. The
$\alpha$-parameter determined for this PN sample is an estimate
of the total number of PNs within 2.5 magnitudes of the bright cut-off
of the PNLF, because of the magnitude limit
of this survey at $m_{5007}$=28.8 and the bright cut-off at
$m^*_{5007}$=26.3 (see Section~\ref{sec5.2}). The measured value of
$\alpha$ is derived from the scaling factor required to match the
PN number density profile to the surface brightness profile in the V
band and then taking the appropriate bolometric
correction into account.

In Fig.~\ref{density_profile}, we showed the total PN number density
profile for the spectroscopically confirmed PNs sample within 2.5
magnitudes from the bright cut-off, together with the halo and IC PN
number densities. The total PN number density profile
flattens at large radii compared to the V-band surface brightness
profile. As already shown in \citet{longobardi13}, this effect
  is due to a superposition of stellar populations with different PN
  specific frequencies. The flattening for the spectroscopocally
  confirmed PN number density profile provides independent support for
  these results. Following \citet{longobardi13}, we can model it by
two components such that
\begin{eqnarray}
\tilde{\sigma}(R)  = \left[\alpha_{\mathrm{halo}}
  \mathrm{I}(R)_{\mathrm{halo,bol}}
  +\alpha_{\mathrm{ICL}}\mathrm{I}(R)_{\mathrm{ICL,bol}}\right] 
\label{mod_sigma}\\
= \alpha_{\mathrm{halo}} \left[\mathrm{I}(R)_{\mathrm{K09,bol}}
    +\left(\frac{\alpha_{\mathrm{ICL}}}{\alpha_{\mathrm{halo}}}-1\right)\mathrm{I}(R)_{\mathrm{ICL,bol}}\right],
\label{mod_sigma_ratio}
\end{eqnarray}
where $\tilde{\sigma}(R)$ represents the predicted total PN surface
density in units of $\mathrm{N}_\mathrm{{PN}}\mathrm{pc^{-2}}$;
$\mathrm{I}(R)_{\mathrm{halo}}$ and $\mathrm{I}(R)_{\mathrm{ICL}}$,
with and without subscript $\mathrm{bol}$, are the bolometric and
V-band surface brightnesses for the halo and the IC components,
in $L_\odot\mathrm{pc}^{-2}$, respectively; $\mathrm{I}_{\mathrm{K09}}$ is the M87
luminosity profile in the V band in $L_\odot\mathrm{pc}^{-2}$, from
\citet{kormendy09}. The surface brightnesses $\mathrm{I}(R)_{\mathrm{halo}}$ and
$\mathrm{I}(R)_{\mathrm{ICL}}$ are given by the Sersic
fit to the observed M87 surface brightness data
\citep[$n=11.8$,][]{kormendy09}, and by the scaled power-law fit to
the IC surface density data from Sect.~\ref{sec4}, respectively.  They satisfy the
relation
$\mathrm{I}_{\mathrm{K09}}=\mathrm{I}(R)_{\mathrm{halo}}+\mathrm{I}(R)_{\mathrm{ICL}}$,
which determines the normalisation of the IC surface brightness
profile.  In the surveyed area over a radial range $7 \mathrm{kpc} < R
< 150 \mathrm{kpc}$, these profiles give total V-band luminosities
$L_{\mathrm{halo}}=4.41\times 10^{10} L_{\odot}$ and
$L_{\mathrm{ICL}}=0.53\times 10^{10} L_{\odot}$, and after the
bolometric correction (see below), total bolometric luminosities for the sampled
halo and ICL of $L_{\mathrm{halo},bol}=9.05\times 10^{10}
L_{\odot,\mathrm{bol}}$ and $L_{\mathrm{ICL},bol}=1.1\times 10^{10}
L_{\odot,\mathrm{bol}}$.

The surface luminosity $\tilde{\sigma}(R)$ can be related to the
bolometric surface brightness through the formula
\begin{equation}
  \tilde{\mu}(R)=-2.5\log_{10}\tilde{\sigma}(R)+\mu_0,
  \label{mod_mu}
\end{equation}
where $\mu_0$ is given by the analytical function
\begin{equation}
  \mu_0=2.5\log_{10}\alpha_{\mathrm{halo}}+\mathrm{K}+(\mathrm{BC_{\odot}-BC_V}).
\label{mu0}
\end{equation}
In Eq.~\ref{mu0}, $\alpha_{\mathrm{halo}}$ is the specific PN number
for the halo, $\mathrm{K}=26.4$ mag arcsec$^{-2}$ is the V-band
conversion factor from mag arcsec$^{-2}$ to physical units
$L_{\odot} \mathrm{pc}^{-2}$, $BC_{\odot}=-0.07$ is the
V-band bolometric correction for the Sun, and $BC_V$=-0.85 is the
bolometric correction for the V-band \citep{buzzoni06}. According to
their simple stellar population (SSP) models for irregular, late, and
early-type galaxies, this value can be used with 10\% accuracy.

From the offset value $\mu_0= 16.56\pm0.08$ mag arcsec$^{-2}$
determined from the density profile in Fig.~\ref{density_profile}, we
compute $\alpha_{\mathrm{halo}} = (1.06 \pm 0.12) \times 10^{-8}$
PN~L$_{\odot,\mathrm{bol}}^{-1}$. From Eqs.~\ref{mod_sigma} and
\ref{mod_sigma_ratio}, the derived value for $\alpha_{\mathrm{ICL}}$
is then $(2.72 \pm 0.72) \times 10^{-8}$
PN~L$_{\odot,\mathrm{bol}}^{-1}$, using the steeper slope -0.34 for
the ICL, but the difference for the shallower slope is only
$5\%$. The difference in $\alpha$-parameters is then
  $\sim2.3 \sigma$.

These luminosity specific PN $\alpha$ values are consistent with those
obtained by \citet{longobardi13} from the photometric sample, and are
now independently validated on the basis of the spectroscopically
confirmed PNs. Moreover, they also agree within the
  uncertainties with the preoviously determined values by
  \citet{durrell02,doherty09}; see \citet[][]{longobardi13}. The
consistency of the spectroscopic and photometric values confirms the
accuracy of the estimated contamination by Ly$\alpha$ background
objects in the photometric sample.

We can now compare our $\alpha_{\mathrm{halo}}$ and
$\alpha_{\mathrm{ICL}}$ values with the known
$\alpha$ values for PN populations in nearby galaxies. Galaxies with
integrated $(B-V)$ colours smaller than $0.8$ are empirically
characterised by similar values of the $\alpha$ parameter,
$\alpha\sim3\times10^{-8}$$\mathrm{N_{\mathrm{PN}}
  L_{\odot,\mathrm{bol}}^{-1}}$, with a scatter of a factor of
two. For redder galaxies with $(B-V) > 0.8 $, the spread of the
measured values increases, spanning a range from
$\alpha \sim10^{-9}$$\mathrm{N_{\mathrm{PN}}/L_{\odot,\mathrm{bol}}}$
to
$\sim6\times10^{-8}$$\mathrm{N_{\mathrm{PN}}/L_{\odot,\mathrm{bol}}}$.
For these redder galaxies, there is an empirical inverse correlation
of the $\alpha$ values with the $FUV-V$ integrated colours of the
parent stellar population, such that smaller $\alpha$ values are
associated with galaxies with larger FUV-V excess. Hence,
observationally, the $\alpha$ values are linked to the metallicity and
star formation history of the parent stellar population of the PNs
\citep{peimbert90,hui93,buzzoni06,longobardi13}.

Our result that the IC component contributes more PNs per unit
bolometric luminosity than the M87 halo light therefore signals a
change in the stellar populations from halo to ICL, consistent with
the existence of a gradient towards bluer colours of the M87 stellar
light at large radii \citep{liu05,rudick10}. We interpret this
gradient as the result of the gradual transition from the redder halo
light to the bluer ICL with decreasing surface brightness.

\subsection{The halo and ICL PNLFs}
\label{sec5.2}
The PNLF of [OIII]$\lambda$5007\AA\ emission line fluxes is often
empirically described via the truncated exponential formula
\citep{ciardullo89}
\begin{equation}
  \mathrm{N}(M) = c_1e^{c_{2}M}\left\{1-e^{3(M^*-M)}\right\},
  \label{PNLF_eq}
\end{equation}
where $c_1$ is a normalisation constant, $c_{2}=0.307$, and
$M^*$(5007)=-4.51 mag is the absolute magnitude of the PNLF bright
cut-off. This analytical formula is designed: (i) to reproduce the high-mass
cut-off observed for the PN central stars in nearby galaxies, and
ii) to model PNs as uniformly expanding homogeneous spheres ionised by
a non-evolving central star \citep{henize63}.

\begin{figure}
  \centering
  \includegraphics[width=8.5cm]{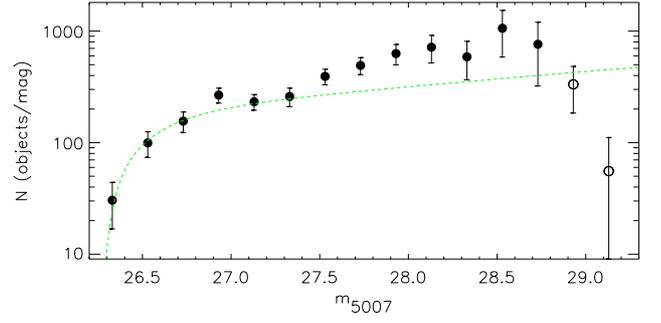}
  \caption{Luminosity function of all spectroscopically confirmed PNs
    corrected for incompleteness (full dots).  The dashed green line
    represents the \citet{ciardullo89} formula for a distance modulus
    of 30.8, convolved with photometric errors and normalised to the
    data at the bright end.  The spectroscopically confirmed PNLF
    shows an excess of fainter PNs with respect to the analytical
    formula, similar to the photometric PNLF
    \citep{longobardi13}. Open circles represent magnitude bins where
    the sample is not completeness-corrected. The error bars
      show the uncertainties from counting statistics and completeness
      correction.}
\label{total_PNLF}
\end{figure}

Our deep and extended imaging survey of PNs in the outer regions of
M87 shows that the PNLF for this galaxy has significant deviations
from Eq.~\ref{PNLF_eq} in the faint magnitude bins
\citep{longobardi13}: its slope $\sim$1-2 magnitudes below the bright
cut-off is steeper than expected from the \citet{ciardullo89}
formula. This is also true for the spectroscopically confirmed PN
sample, as we show in Fig.~\ref{total_PNLF} in which we compare the total
PNLF with the analytical formula for a distance modulus of 30.8.

\noindent
\begin{figure}
  \centering
  \includegraphics[width=8.5cm]{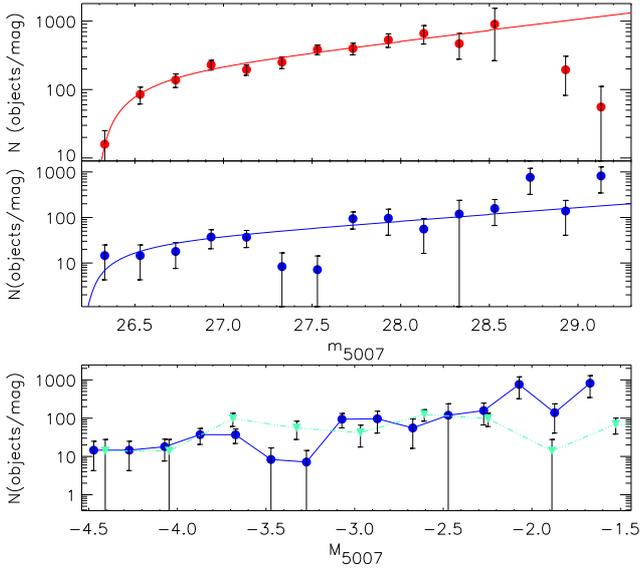}
  \caption{\textit{Top panel}:~Luminosity function of the
    spectroscopically confirmed halo PNs, corrected for incompleteness
    (red circles).  The red line shows the fit of the generalised
    analytical formula (Eq.~\ref{PNLF_eq}) to the halo PNLF, with
    $c_2=0.72$ and bright cut-off at 26.3 mag, corresponding to a
    distance modulus $m-M=30.8$.
    \textit{Central~panel}:~Completeness-corrected PNLF for the ICPNs
    (blue circles). The blue line shows the fit of the generalised
    analytical formula (Eq.~\ref{PNLF_eq}) to the ICPNLF, with
    $c_{2}=0.66$. The ICPNLF shows a dip 1-1.5 mag fainter than the
    bright cut-off.  \textit{Bottom~panel}:~PNLFs in absolute magnitude
    for the spectroscopically confirmed Virgo ICPNs (blue circles) and
    for the outer disk of M33 \citep[][cyan triangles]{ciardullo04}
    matched at the first bin.  For this comparison, the data are
    corrected for line-of-sight reddening (see text).  Absolute
    magnitudes are determined using a distance modulus of 30.8 for the
    ICL and 24.86 for M33 \citep{ciardullo04}.}
\label{PNLF}
\end{figure}

Our data allow us to analyse separately and compare the PNLFs of the
two spectroscopically confirmed PN samples for the M87 halo and ICL,
from the bright cut-off down to 2.5 mag below.  In the
upper panel of Fig.~\ref{PNLF}, we show the PNLF for the
spectroscopically confirmed sample of halo PNs, corrected for
detection incompleteness as a function of magnitude.  The data points
trace a smooth function, and the fit of the generalised analytical
formula (Eq.~\ref{PNLF_eq}) to the observed PNLF within the 2.5 mag
limit results in $c_{2}=0.72$ and a bright cut-off at $m_{5007}^{*}=
26.3$ (overplotted on the data). With $M^{*}=-4.51$, this corresponds
to a distance modulus ($m-M$)=30.8.

In the central panel of Fig.~\ref{PNLF}, we present the ICPNLF (full
blue circles), corrected for incompleteness. As for the M87 halo PNLF,
the ICPNLF is consistent with the same distance modulus as for
M87. However, it shows an overall shallower gradient than the M87 halo
PNLF: the best fit of Eq.~\ref{PNLF_eq} to the empirical PNLF returns
$c_2=0.66$. 

In addition to the shallower gradient at fainter magnitudes, the
ICPNLF shows a clear dip at 1-1.5 mag fainter than
$m_{5007}^{*}$. This feature is statistically significant: the
difference between the number of PNs in these bins with respect to the
magnitude bins before and after the dip is
$>\,3\,\sigma_{\mathrm{dip}}$ on both sides, where
$\sigma_{\mathrm{dip}}$ is the uncertainty from Poisson statistics in
the magnitude bins where the dip occurs.

Dips in the PNLF are observed for PN populations detected in star-forming
galaxies (irregulars/disks), and are absent in the PNLFs of
bulges or early-type galaxies. The magnitude below the bright cut-off
at which the dip occurs varies between different PN populations, from
$\sim2$ to $\sim4$ mag below $M^{*}$
\citep{jacoby02,ciardullo04,hernandez09,reid10}. We discuss this issue
further in Section~\ref{sec6.1}.  In the bottom panel of
Fig.~\ref{PNLF}, we compare the PNLFs for the Virgo ICPNs and for the
spectroscopically confirmed sample of PNs for the outer disk of M33
\citep{ciardullo04}. Both LFs are corrected for foreground Galactic
extinction, adopting reddening values of $E(B-V)_{\mathrm{ICL}}$=0.02
\citep{ciardullo98}, and $E(B-V)_{\mathrm{M33}}$=0.04
\citep{ciardullo04} for the Virgo ICL and M33,
respectively\footnote{We used the relation of \citet{cardelli89} with
  R$_{V}=3.1$ to go from reddening to extinction at
  5007\AA.}. Absolute magnitudes are determined using a distance
modulus of 30.8 for the ICL and 24.86 for M33
\citep{ciardullo04}. Both PNLFs show dips relative to the smooth
luminosity function: the ICL at 1-1.5 mag fainter than $M^{*}$, the
outer disk of M33 at $\sim2.5$ mag fainter than the bright cut-off.

Finally, we recall that about 17\% of the PNs contained in the M87
halo PNLF shown in Fig.~\ref{PNLF} are ICPNs whose velocities fall in
the same velocity range of the M87 halo PNs, and can therefore not be
individually identified. The hint of a slight dip in the halo PNLF at
$\sim1$ mag below the bright cut-off may be due to these ICPNs.

To summarise, the observed properties of the M87 halo and Virgo ICPN
populations i.e. their $\alpha$-parameters and PNLFs, show
significant differences. Because these quantities depend on the
physical properties of the parent stellar populations, these
differences imply that the M87 halo and ICL consist of different
populations of stars.  To understand this better, more work is clearly
required for a better theoretical understanding of how metallicity,
age, and different star formation histories affect the post-AGB phases
of stellar evolution and the resulting PN populations.

 \section{Discussion} 
\label{sec6}

\subsection{The distinct halo and IC populations around M87}
\label{sec6.1}

In Sect.~\ref{sec3} we presented the projected phase-space
distribution of the spectroscopically confirmed PNs in our M87 fields
(Fig.~\ref{Pspace}). With a robust procedure we showed that the PN
velocity distribution splits into two kinematically very different
components: the M87 halo (with mean velocity V$_{\mathrm{LOS,n}}
= 1275$ kms$^{-1}$ and velocity dispersion $\sigma_{\mathrm{n}}
\simeq 300$ kms$^{-1}$) and the ICL (with V$_{\mathrm{LOS,b}} \simeq
1000$ kms$^{-1}$ and $\sigma_{\mathrm{b}} \simeq 900$ kms$^{-1}$). Furthermore,
in Sect.~\ref{sec5}, we found that the halo and IC components
were characterised by specific PN numbers ($\alpha$ parameters) that
differed by a factor of three, and by different shapes of their PNLFs.

These results demonstrate the coexistence of two distinct PN
progenitor stellar populations in this region of the Virgo cluster
core: the M87 halo and the ICL. These two populations have very
different surface density distributions, the M87 halo is described
by an $n=11.8$ Sersic law, while the ICL follows a shallow power law
$\propto R^{\gamma}$ with $\gamma$ in the range $[-0.34,-0.04]$. We
also have external information on the metallicities and ages of both
components. At $R\sim 35$kpc, the mean metallicity of the M87 halo
obtained with population synthesis models from multi-colour photometry
is $\sim 0.7$ solar, with a shallow outward gradient, and the mean age
is $\sim 10$Gyr \citep{liu05, Montes14}. On the other hand, the
metallicity and age distributions of ICL red giants in a field at
$R\sim 190$ kpc from HST ACS star photometry are dominated by
metal-poor ([M/H]$\lta -1$), $\gta 10$Gyr old stars
\citep{williams07}. Because of the large velocities and shallow
surface density profile of the IC stars, these IC population
parameters are likely to be similar in the radial range probed by our
observations, $R\sim50-140$ kpc, whereas the M87 halo stars might reach
$\sim 0.5$ solar in the outer regions if the outward gradient
continues, as inferred by \citet{liu05}.

Currently there is no good theoretical understanding of how the
properties of a PN population are related to the metallicity and age
of a stellar population. Observationally, star-forming and bulge
populations have $\alpha$ numbers such as those we find for the ICL, while
only the most massive early-type galaxies have $\alpha$ numbers as low
as we find for the M87 halo \citep{buzzoni06,cortesi13}. The primary
driver is believed to be increased mass loss at high metallicities.
The PNLFs are empirically found to steepen from star-forming to old
metal-rich populations \citep{ciardullo04,longobardi13}. The PNLFs of
Local Group star-forming galaxies, such as the SMC \citep{jacoby02},
LMC \citep{reid10}, M33 \citep{ciardullo04}, and NGC 6822
\citep{hernandez09}, furthermore show a `dip' 2-2.5 mag down from the
PNLF cut-off for the LMC, M33, and NGC 6822, and 4 mag down from the
cut-off for the SMC. A tentative model for this feature is the
superposition of a faint PN population with a brighter population of
more massive cores from a younger stellar population
\citep{rodgonz2014}. We can speculate that as the brighter population
fades in older and/or more metal-rich populations, the dip might move
towards brighter magnitudes.  This could explain why in the Virgo IC
population we find the dip 1-1.5 mag down from the cutoff. No other PN
population with this PNLF is known; however, PNLFs as deep as for M87
have only been obtained in the Local Group so far. Clearly, more
observational and theoretical work on the nature and location of the
dip in the PNLF is needed.

\subsection{The ICL in Virgo: nature and number of its progenitor
  galaxies}
\label{sec6.2}

The combined properties of the Virgo ICPN population, which include the fairly
small inferred bolometric luminosity, the relatively large $\alpha$
parameter, and the dip in the PNLF as well as the low mean
metallicity from \citet{williams07}, appear to be most readily
explained if this population derives from a faded population of
low-luminosity, low-metallicity, star-forming or irregular galaxies,
such as M33 or the LMC, which are very different from M87 itself.

In Section~\ref{sec5.1}, we determined the total V-band and bolometric
luminosities of the IC component sampled by our survey fields:
$L_{\mathrm{ICL}}=0.53\times 10^{10} L_{\odot}$, and
$L_{\mathrm{ICL,bol}}=1.1\times 10^{10} L_{\odot,\mathrm{bol}}$.
Using the total V-band luminosities for M33 and the LMC listed in NED,
$L_{\mathrm{M33}}$=3.65$\times 10^{9} L_{\mathrm{\odot}}$ and
$L_{\mathrm{LMC}}$=1.26$\times 10^{9} L_{\mathrm{\odot}}$, we find
that the IC stars sampled in our survey fields correspond to $\sim
1.5$ M33-like galaxies or $\sim 4$ LMC-like galaxies\footnote{The total
luminosity at all radii corresponding to the detected IC
  stars is much larger; their large measured velocity dispersion
  implies that the orbits of these IC stars reach to much larger
  radii in the cluster.}.

We can now also check whether the Virgo ICL associated with the ICPN
population could be related to the blue GC population that is found
around M87, that has a shallower and more extended surface density profile
than the red GCs, which trace the stellar halo light
\citep{cote01,tamura06,strader11,forte12,durrell14}.  To do this, we
need to estimate the total number of GCs associated with M33 and
LMC-like systems. \citet{harris13} studied GC populations in a large
sample of galaxies and analysed the correlation of the total number of
GCs, $N_{\mathrm{GC}}$, with global galaxy properties and type. They
find that $N_{\mathrm{GC}}$ increases roughly in direct proportion to
host galaxy luminosity, with a scatter of a factor of $\sim 2.5$. For
an LMC-like system with luminosity L$_{\mathrm{LMC}}$=1.3$\times
10^{9}$L$_{\mathrm{\odot}}$ the expected mean number of GCs is
$N_{\mathrm{GC}}\sim20$, while for an M33-like galaxy with
L$_{\mathrm{M33}}$=3.7$\times 10^{9}$L$_{\mathrm{\odot}}$ the expected
mean number of GCs is $N_{\mathrm{GC}}\sim60$.  This leads to an
estimated number of blue GCs associated with the sampled ICL of
$N_{\mathrm{GC,ICL}}\sim80-90$, with a scatter of a factor $\sim 2.5$.
If a fraction of the ICL is due to the accretion of even lower
luminosity galaxies, the estimated number of GCs would increase
\citep{harris13,coccato13}.

The recent survey of the GC population in the Virgo cluster around M87
by \citet{durrell14} showed the presence of an ICGC population,
mostly associated with blue GCs (see further discussion in
Sect.~\ref{sec6.3}). This intracluster component has a density equal
to $\Sigma_{bGC,tot}=0.2^{+0.13}_{-0.08}$arcmin$^{-2}$. In our
surveyed region, this would lead to a total number of 100-430 ICGCs.
This is larger than but consistent within the uncertainties with the
value estimated above, suggesting that a substantial fraction of the
blue GC population around M87 could have been accreted with the
galaxies that we now see in the ICL.

\subsection{Is there an intracluster component of globular clusters
  around M87?}
\label{sec6.3}

There has been some controversy in the recent literature about the
existence of intracluster GCs in the halo of M87. The most extensive
photometric study of the distribution of GCs in the Virgo cluster so
far was carried out by \citet{durrell14} as part of the NGVS.  They
studied density maps of the GC population, selected using colour
criteria, and statistically accounted for the contamination to the GC
sample by subtracting a modelled map for the expected background, from
both Milky Way stars and background galaxies\footnote{We note here
  that the M87 PN sample cannot be contaminated by Milky Way halo PNs,
  as these would have [OIII]$\lambda$5007\AA\ fluxes about 12 mag
  brighter than M87 PNs.}.  The blue GCs in their map have a shallower
and more extended profile than the red GCs.  \citet{durrell14}
  also found that the total GC (blue plus red) density profile is in
  good agreement with the number density profile of photometrically
  selected PNs from \citet{longobardi13}, including a change of slope
  and a flatter profile at large radii. They suggested that their
blue GCs at distances $> 215$ kpc are part of the intracluster
component of Virgo. Cosmological simulations
  \citep{dolag10,cui14,cooper14} predict that the density of this
  component would then increase inwards. 

\citet{durrell14} also found evidence for a spatial asymmetry of GCs
surrounding M87 for major axis distances larger than 20$\arcmin$, with
an excess of tracers in the NW region (mostly the blue population). In
Sect.~\ref{sec4} we studied the distribution of M87 halo and IC PNs
separately for the NW and SE.  We find no clear evidence of asymmetry
in either the halo and ICL within major axis distance $\sim20\arcmin$
(Fig.~\ref{density_profile_NS}).  Inside this radius, both PN and GC
number density profiles are consistent with a symmetric halo and IC
distribution.  For the halo component, this result is significant,
given the number of tracers and radial extent, and indicates, that if
the halo was subject to accretion events, these were not
recent. For the ICL, we may expect asymmetries, given the longer
timescales involved in IC accretion events, but we may not have a
large enough sample of ICPNs to see them.

\citet{strader11} carried out a spectroscopic study of the GCs around
M87, using colour criteria to select their candidates.  In the same
colour and magnitude range populated by globular clusters, $0.55 \le
(g'-i') \le 1.15$ and $20 \le g' \le 24$, there is however a large
contribution from foreground Milky Way halo stars. To mitigate
  this effect, \citet{strader11} used a combination of
  photometry, radial velocity, and HST imaging information. However,
  they considered all objects with velocity $\mathrm{V_{LOS}} < 150\,
  \rm{kms^{-1}}$ to be stars. In the most ambiguous range, i.e.
  $\mathrm{150 \rm{kms^{-1}} < V_{LOS} < 350} \rm{kms^{-1}}$, they
  classified all uncertain objects, for which a clean separation
  between contaminants and GCs could not have been done, as stars.
Based on the remaining sample, they reported that the
number density profile of the spectroscopically confirmed GCs showed
no evidence for a transition between a halo and IC component, either
as a sharp truncation of the halo, or a flattening of the GC number
density profile at large radii. From the sample kinematics, they
observed that their GCs around M87 have velocity dispersion in the
range $300\le\sigma\le500$ kms$^{-1}$ out to 190 kpc, with $\sim 500$
kms$^{-1}$ for the GCs population at 190 kpc significantly smaller
than the velocity dispersion of Virgo cluster galaxies
\citep{binggeli93,conselice01}. However, from the PN phase-space distribution
($\mathrm{V_{LOS,PN}}, R_{\mathrm{PN}}$) in Fig.~\ref{Pspace}, we see
that a large fraction of IC stars near M87 have velocities
$\mathrm{V}_{\mathrm{LOS}}<350$ kms$^{-1}$. This suggests that the
lack of evidence for the IC component reported by \citet{strader11}
could be caused by the velocity threshold $\mathrm{V_{min}} = 350$
kms$^{-1}$ imposed on the GC sample, which is needed to prevent the
contamination from Milky Way halo stars, but may also remove many of
the ICGCs from their analysis.

\subsection{Relation between BCG and ICL}
When studying central galaxies in clusters, one of the main question
is to establish where the ICL begins and where the associated BCG
ends, or whether any distinction is to be made at all.  

For M87, the differences in the density profiles and velocity
distributions of the halo and ICPN populations, as well as in their
$\alpha$-parameters and PNLFs, are sufficient to argue that the two
components must be considered to be separate stellar populations, with
different metallicities and star formation histories, and not as a
continuum.  As discussed above, published stellar population data
suggest that the halo stars are older and more metal-rich than the
ICL (see Sects.~\ref{sec5}, \ref{sec6.1} for more details).

In more distant BCGs where a kinematic decomposition between BCG halo
and ICL is not available, the presence of an additional dynamical
component in BCGs is usually inferred from a change of slope at large
radii in the SB profile \citep{zibetti05,gonzales07,dsouza14}.
Photometric properties or colours are obtained by treating the two
components as a continuum because no differentiation between the
underlying stellar populations is normally possible.

Using a particle tagging method to analyse galaxy clusters in
$\Lambda$CDM simulations, \citet{cooper14} consider the BCG and ICL as
a single entity consisting of all stars, which are not bound to any
cluster subhalos. They then split the BCG stars into accreted
  stars and in situ stars, and find that the large majority
of BCG stars are accreted stars.  They find double-S\`ersic surface
density profiles in their simulated BCGs, where the inner component
($R<200 {\rm kpc}$) is dominated by `relaxed' accreted components, and
the outer component by `unrelaxed' accreted components.
\citet{cooper14} argue that the accreted/in situ separation is
physically meaningful and that the ICL should naturaly be considered
as a continuation of the BCG to low surface brightness because both
components are formed by similar mechanisms.

In contrast, a dynamical approach based on the velocity distributions
of diffuse light particles in hydrodynamic cosmological simulations
\citep{dolag10,cui14} is found to separate these stars into two
components: one bound to the cluster potential and the other bound to
the BCG. The resulting BCG and diffuse ICL are formed on
different timescales, and the simulated stars associated with the two
components are different in terms of spatial distribution, ages, and
metallicities.  \citet{cui14} showed that it is possible to
dynamically differentiate between halo and IC particles using the
particles' binding energies. Stars with high binding energy, which end
up belonging to the BCG, were subjected to relaxation and merging
processes such that the gravitational potential changed so quickly
that these stars lost memory of the kinematics of their progenitors
\citep{murante07, dolag10}. On the other hand, stars with lower
binding energy, that belong to the diffuse component, still reflect
the dynamics of the satellite galaxies. Both \citet{dolag10} and
\citet{cui14} also observed that the slope of the surface brightness
profile associated with the two components change, with the halo
profile being steeper than the ICL profile.

It is likely that the distinct BCG and IC components found in the
hydrodynamical simulations are related to the relaxed and unrelaxed
accreted components in the particle tagging analysis, but the
inclusion of baryonic processes in the former may accentuate the
differences found between BCG and ICL.  If we associate the BCG and
ICL of \citet{dolag10, cui14} with the relaxed and unrelaxed accreted
components of \citet{cooper14}, the progenitors of the stars in the
steeper S\`ersic (relaxed) component would be accreted from more
massive systems at higher redshifts, while the stars in the shallower
and more extended ICL (unrelaxed) component would come from the
accretion of less massive systems at lower redshifts. More massive
progenitors dominate the diffuse light in simulated clusters close to
the centre \citep{murante07, puchwein10} because they move further inwards 
by dynamical friction. They cause stronger relaxation of the
gravitational potential, and if accreted early, they have more time to
relax.

To summarise, recent simulations show that a distinction can be made
between stars that trace the cluster potential and stars bound to the
BCG, based on the physical properties and binding energies of the
accreted progenitors.  From the study of the PN population around M87,
we have shown the coexistence of two discrete components in the Virgo
cluster core, tracing different stellar populations, in agreement with
these predictions. While the PN population for the IC component
around M87 indicates low-mass dwarf and star-forming galaxy
progenitors, the stellar halo has higher metallicity, $\sim 0.7$
solar, indicating more massive progenitors. This bimodality in the
progenitors may be the root of the bimodality in the kinematics and
density profiles of the M87 halo and the ICL.

However, we note that this kind of bimodality need not occur in every
cluster of galaxies. For example, it is plausible that, for a more
continuous distribution of progenitor masses and a more uniform
distribution of binding energies of the debris stars, the final BCG
plus ICL system would show continuous radial gradients in kinematics
and stellar population properties, rather than appear as the sum of
several discrete components. It is possible that NGC 6166 in the Abell
2199 cluster is closer to this situation: the velocity dispersion in
the high surface brightness halo of this BCG was recently measured to
increase up to the cluster velocity dispersion of $\sim 800 {\rm
  km}s^{-1}$ at 100'' from the galaxy centre (\citet{bender14}; see
also \citet{kelson02}).

 \section{Summary and conclusions} 
\label{sec7}

We obtained spectra for 287 PNs in the outer regions of the nearby
elliptical galaxy M87, of which 211 are located between distances 40
kpc to 150 kpc from the galaxy centre.  Spectra were acquired with the
FLAMES spectrograph in the GIRAFFE+MEDUSA configuration, with spectral
resolution of $R=22500$. We observed 14 different FLAMES plate
configurations, using candidates from the catalogue described in
\citet{longobardi13}. The spectroscopic survey aimed at measuring the
LOS velocities of PNs in the transition region between the galaxy's
stellar halo and the ICL. We identified PNs through their narrow and
symmetric, redshifted [OIII]$\lambda$5007\AA\ emission line, with no
or negligible continuum, and verified with the second
[OIII]$\lambda$4959\AA\ emission line. Spectra were measured for PNs
in the magnitude range from m$_{5007}$=26.3 down to 28.8. This is the
largest spectroscopic sample of PNs at such galactic radii for a
central galaxy, in the number of tracers and magnitude depth.

The area covered by the survey allowed us to trace the transition
between the M87 halo and ICL in the Virgo cluster core. The
  coexistence of these two components is shown by the bimodality of
  the LOSVD, whose strong asymmetric wings make it deviate from the
  near-Gaussian LOSVD typical of early-type galaxies. We separated
halo and ICPNs by studying the projected phase-space distribution.
We implemented a robust technique to measure the velocity dispersion
of the M87 halo, separating its velocity distribution from the broader
component, the ICL. We identified 243 PNs for the M87 halo and 44
ICPNs.  We found that the logarithmic number density profile for the
halo PNs follows the V-band SB profile from \citet{kormendy09}, while
the IC number density profile decreases towards large radii as a
power-law $I_{\mathrm{ICL}}\propto\mathrm{R}^{\gamma}$ with $\gamma$
in the range $[-0.34,-0.04]$.

The total PN surface density profile at large radii is flatter than
the surface brightness profile because of the presence of the IC
component \citep[see also][]{longobardi13}, which contributes $\sim$3
times more PNs per unit luminosity than the halo population.  We find
luminosity-specific PN numbers
$\alpha_{\mathrm{halo}}=(1.06\pm0.12)\times 10^{-8}$
N$_{\mathrm{PN}}L^{-1}_{\odot,\mathrm{bol}}$ and $
\alpha_{\mathrm{ICL}}=(2.72\pm0.63)\times10^{-8}$
N$_{\mathrm{PN}}L^{-1}_{\odot,\mathrm{bol}}$ for the M87 halo and IC
PN population, respectively. This is consistent with the known
existence of a gradient towards bluer colours at large radii because of
the increased contribution of ICL at large distances and its lower
metallicity compared to the halo population.

The spectroscopically confirmed PNLFs for both the halo and IC PNs
have a steeper slope towards faint magnitudes than is predicted by the
analytical formula of \citet{ciardullo89}, confirming the result from
the photometric sample \citep{longobardi13}. This steepening is
consistent with an old stellar population dominated by PNs with low-mass
cores.  The PNLF of the ICPN population has a slightly shallower
gradient than the M87 halo PNLF, and in addition shows dip at about
$\sim$ 1 - 1.5 magnitudes from the bright cut-off. This dip is an
evolutionary feature observed in star-forming systems, such as M33 and
the Magellanic clouds, and may be related to rapidly evolving PNs with
massive central cores. The presence of the dip in the ICPNLF, but not
in the M87 halo PNLF, provides additional evidence for intrinsic
differences between the halo and IC parent stellar populations.

Using PNs as tracers we showed that the stellar halo of
the BCG galaxy M87 is distinct from the surrounding ICL in its
kinematics, density profile, and parent stellar population, consistent
with the halo of M87 being redder and more metal-rich than the ICL.
We note that the ICL in our surveyed fields corresponds to about four
times the luminosity of the LMC, spread out over a region of $\sim
100\, {\rm kpc}$ diameter. It is remarkable that population properties
can be observed for such a diffuse component.

In the Virgo cluster, BCG halo and ICL cannot be considered as
components with a gradual transition in their kinematics. This
supports results from analysis of galaxy cluster simulations, which
suggest that the IC component in Virgo consists of unrelaxed accreted
stars bound to the cluster potential, while the stellar halo of M87
appears to be described as a relaxed accreted component bound to the
galaxy itself.  Based on its PN population properties, we propose that
the progenitors of the Virgo ICL were low-mass, star-forming galaxies,
which may also have brought with them a significant fraction of the
blue GC population seen in the outer regions of M87.

%\newpage

\begin{acknowledgements}
  We thank Ken Freeman, John Kormendy, and Sadanori Okamura for
  discussions and helpful suggestions, and the ESO user support
  department and the La Silla - Paranal science operation staff for
  the support of our service mode observations. AL is grateful to
  J.~Elliott, and B.~Agarwal for helpful discussions. This work made
  use of the NASA/IPAC extragalactic database (NED) operated by the
  Jet Propulsion Laboratory and the California Institute of
  Technology.
\end{acknowledgements}

\bibliographystyle{aa}
\bibliography{PNrefs}

\end{document}